\documentclass[12pt]{article}

\usepackage{mathrsfs,amssymb,amstext,amsmath,amsthm,eucal}
\usepackage{graphicx,subfig}
\usepackage{authblk}
\usepackage[usenames,dvipsnames]{color}
\usepackage{tikz}\usetikzlibrary{arrows,shapes}
\usepackage{geometry}
\usepackage{mcite}
\usepackage{booktabs}
\usepackage{hyperref}

\usepackage{pdflscape,afterpage}

\newcommand{\grad}{\mathrm{d}}
\newcommand{\CC}{\mathbb{C}}\newcommand{\RR}{\mathbb{R}}\newcommand{\ZZ}{\mathbb{Z}}
\newcommand{\dvol}{\mathrm{dvol}}
\DeclareMathOperator{\proj}{proj}
\DeclareMathOperator{\sign}{sign}

\newcommand{\famdeforms}{{\circ\mspace{-3mu}-}}
\newcommand{\circright}{{{\circ}\mspace{-7.5mu}\rightarrow}}
\newcommand{\implodes}{\circright}

\usepackage{enumitem}
\newlist{ECsols}{enumerate}{2}
\setlist[ECsols,1]{label=(EC\arabic{ECsolsi}),resume,leftmargin=*}
\setlist[ECsols,2]{label={(\arabic{ECsolsi}\roman{ECsolsii}.)},leftmargin=*}

\newcommand{\mysosmall}{ iso }
\newcommand{\mysocapital}{ ISO }

\newenvironment{definition}[1][Definition]{\begin{trivlist}
\item[\hskip \labelsep {\bfseries #1}]}{\end{trivlist}}

\author{George Moutsopoulos\footnote{gmoutso@gmail.com}}
\title{Homogeneous anisotropic solutions of topologically massive gravity with cosmological constant and their homogeneous deformations}
\affil{Department of Electrophysics, National Chiao-Tung University, Hsinchu, Taiwan}
\affil{Riemann Center for Geometry and Physics, Leibniz University Hannover, Hannover, Germany}
\date{\today}

\begin{document}
\maketitle
\begin{abstract}
We solve the equations of topologically massive gravity (TMG) with potentially non-vanishing cosmological constant for homogeneous metrics without isotropy. We only reproduce known solutions. We also discuss their homogeneous deformations, possibly with isotropy. We show that de Sitter space and hyperbolic space cannot be infinitesimally homogeneously deformed in TMG. We clarify some of their Segre-Petrov types and discuss the warped de Sitter spacetime.
\end{abstract}

\newpage
\tableofcontents
\newpage

\section{Introduction}
We study two related questions about three-dimensional topologically massive gravity (TMG) \cite{Deser:1981wh,Deser:1982vy}. In the first, we ask what are the homogeneous anisotropic solutions of TMG with a generically non-vanishing cosmological constant. That is, we solve the equations of motion of topologically massive gravity with a cosmological constant on three-dimensional Lie groups that possess a left-invariant metric. In the second, we ask what are their homogeneous deformations. That is, we ask how one can continuously deform the metric of a homogeneous anisotropic solution such that the spacetime remains both a solution and a homogeneous space, possibly with isotropy. The three-dimensional maximally symmetric spaces solve the equations of TMG. In the second question, we are interested in particular in the homogeneous deformations of maximally symmetric spaces that still solve TMG.

The equations of motion for TMG with cosmological constant $\Lambda$ are that the scalar curvature is constant with $R=\Lambda/6$ and the traceless Einstein tensor is proportional to the Cotton tensor,
\begin{equation}\label{eq:EinCot}
R_{ab} - \frac{R}{3} g_{ab} = \mu\, C_{ab} ~.
\end{equation}
The symmetric traceless Cotton tensor $C_{ab}$ depends on the metric $g_{ab}$ and a  choice of sign for the orientation $\epsilon_{abc}$,
\begin{equation}\label{eq:Cotton}
C_{ab} = \epsilon_{a}{}^{cd}\, \nabla_d \left(R_{bc} - \frac{1}{4} R \,g_{bc}\right) ~.
\end{equation}
The Cotton tensor is identically zero if and only if the space is conformally flat, and so the Einstein spaces, which in three dimensions are locally maximally symmetric, always solve the theory. In TMG we take a constant $R$ and the second term in \eqref{eq:Cotton} is absent, $C_{ab}= \epsilon_{a}{}^{cd} \nabla_d\,R_{bc}$. 

The survey for homogeneous anisotropic solutions of TMG with vanishing cosmological constant, equivalently for $R=0$, was conducted in the late 80s by \cite{Ortiz:1989vc} and \cite{Nutku:1989qi}, and with a comment for isotropic homogeneous spacetimes in \cite{Ortiz:1990nn}. In \cite{Nutku:1993eb}, Y.~Nutku generalizes two particular solutions for the non-vanishing cosmological constant case: solutions with the pp-wave metric Ansatz and what has recently been called warped anti-de Sitter. More recently, \cite{Bakas:2010kc} solve TMG and its generalization of generalized massive gravity with $R\neq0$ on $\mathrm{SU}(2)$ and thus obtain some of the solutions on $\mathrm{SL}(2,\RR)$ by analytic continuation.  The aim of the first part of our paper is to systematically solve TMG for all metrics on all three-dimensional Lie groups.

In general, various solution techniques have resulted to a small class of ubiquitous $R\neq0$ solutions. Let us mention here the supersymmetric solutions, which give a branch of the TMG pp-waves~\cite{Gibbons:2008vi}, and the stationary axisymmetric solutions of \cite{Clement:1994sb,Ertl:2010dh}, whereby the equation \eqref{eq:EinCot} reduces to a one-dimensional conformal mechanics problem; see also \cite{Deser:2009er,Gurses:2008wu,Gurses:2010sm,perfect,Moutsopoulos:2011ez}. The black hole solutions of \cite{Clement:1994sb,Bouchareb:2007yx,Anninos:2009jt} are locally diffeomorphic to a homogeneous deformation of anti-de Sitter space, the so-called warped anti-de Sitter.

The ubiquitous solutions of TMG are warped anti-de Sitter and the TMG pp-waves. The stretched / squashed sphere also satisfies the equation of motion of TMG. When the cosmological constant vanishes, $R=0$, the homogeneous anisotropic solutions are given in the works of \cite{Ortiz:1989vc} and \cite{Nutku:1989qi}. There are a few more solutions in \cite{Ortiz:1989vc,Nutku:1989qi} that are not the special $R=0$ cases of warped anti-de Sitter, pp-waves, or the stretched / squashed sphere. The situation changes with the general Kundt solutions in \cite{Chow:2009vt}. An eloquent description and classification of solutions of TMG is given in \cite{Chow:2009km}, which along with the Kundt spacetimes encompass all known ones. 

We relate one of the Ortiz $R=0$ solutions~\cite{Ortiz:1989vc}, our \ref{item:sl2triaxial}, to the type II constant scalar curvature Kundt. A three-dimensional CSI spacetime is either Kundt or homogeneous~\cite{Coley:2007ib}, or both. We thus highlight a non-trivial homogeneous anisotropic solution that is also Kundt. We will also comment how the $R=0$ homogeneous solutions can be of spacetime type I$_\RR$, I$_\CC$ or II. In this regard, we disagree with the classification of \cite{Chow:2009km} that cite only a solution of type I$_\RR$. However, we do not find any homogeneous anisotropic solution that {is not} a special case of the known solutions. 

Our methodology is quite straightforward. We fix a Lie algebra basis $\tau_a$ up to the automorphism group for {each} Lie algebra $\mathfrak{g}$ of the three-dimensional Lie algebras. The metric $g_{ab}$ at the identity of the group is identified with a metric $B_{ab}$ on the Lie algebra. We then fix the metric $B_{ab}$ on $\mathfrak{g}$ in the space of orbits of metrics under the action of the algebra's automorphisms. The indices $a,b,\ldots$ that appear in \eqref{eq:EinCot} will always refer to a left-invariant frame rather than an orthonormal frame. The equations of motion then reduce to algebraic equations on $B_{ab}$. Our methodology differs from the more common approach that is to fix an orthonormal frame up to local Lorentz or euclidean rotations.

In the second part of the paper we consider the homogeneous deformations of the solutions we obtained. A homogeneous spacetime is given by a coset $G/H$ and an $H$-invariant metric $B$ on the vector space $\mathfrak{g}/\mathfrak{h}$, where $\mathfrak{g}$ and $\mathfrak{h}$ are the Lie algebras of $G$ and $H$ respectively. Although we have solved previously for all $G/H$ TMG solutions with $H=1$, a particular solution might have a larger, enhanced, isometry. In fact, we show in the first part that this happens only for the maximally symmetric spaces, the type D solutions, and the null warped anti-de Sitter. We want to consider continuous deformations $(G_t/H_t,B_t)$, such that at $t=0$ we recover the undeformed spacetime. That is, a homogeneous spacetime can be deformed in its group structure or in the metric at a fixed point of the space. 

At $t=0$ the group $G_0$ should be a subgroup of the undeformed isometry group and such that it acts transitively on the undeformed space. For instance, if we ask for the homogeneous deformations of a maximally symmetric space, say for definiteness anti-de Sitter, we should first list all such transitive subgroups in $\mathrm{SO}(2,2)$. For each such subgroup $G_0$ we then study its infinitesimal deformations. This definition gives rise to the notion of infinitesimal homogeneous deformations. We can also consider the contraction of the three-dimensional Lie algebras. A Lie algebra contraction is a limit that is in a sense opposite to that of an infinitesimal deformation. Although we will consider three-dimensional Lie algebra contractions, we will neglect the abelian contraction overall that gives flat space.

The motivation for this paper was set in motion from a typographical slip\footnote{this appeared in the first version of the arXiv preprint, not in the published version.} in \cite{Anninos:2011vd}, in which a particular solution that is called there warped de Sitter was described as a deformation of three-dimensional de Sitter space. The metric has isometry ${\mathfrak{sl}_2}\oplus\RR$ for all values of the deformation parameter $t$ and ${\mathfrak{sl}_2}$ acts transitively. The metric is what we call warped anti-de Sitter: it is a metric on $\mathrm{SL}(2,\RR)$ such that it is left invariant by the left action of $\mathrm{SL}(2,\RR)$ and a one-dimensional subgroup of the right action. The isometry algebra of de Sitter space is $\mathfrak{sl}(2,\CC)$. Since ${\mathfrak{sl}_2}\oplus\RR$ is not a subalgebra of $\mathfrak{sl}(2,\CC)$, the metric cannot be a deformation of de Sitter or be de Sitter space at any value of $t$. 

We were thus curious as to whether de Sitter space can be homogeneously deformed in the solution space of TMG, just like anti-de Sitter, the sphere, and flat space are known to allow $\mu\neq0$ deformations in \eqref{eq:EinCot}. The subalgebras of $\mathfrak{sl}(2,\CC)$ of dimension at least three that also act transitively on de Sitter space are the four-dimensional Borel subalgebra and a three-dimensional family of subalgebras. For all cases where $(G_0/H_0,B_0)$ is de Sitter, the space cannot be infinitesimally deformed to a $\mu\neq0$ solution, either because $G_0$ is rigid under deformations as in the case of the Borel subalgebra, or because there is a unique fixed solution as in the case of the infinitesimal deformations of the three-dimensional transitive subalgebras. The only possibility is for de Sitter space to contract to flat spacetime.

At another front, the authors of \cite{Gupta:2009zt} analyze the euclidean continuation of warped anti-de Sitter. However, they do not write an explicit metric. The euclidean continuation of anti-de Sitter is hyperbolic space, with isometry $\mathfrak{sl}(2,\CC)$, and hyperbolic space is a solution of TMG by virtue of it being an Einstein space. We show that hyperbolic space cannot be continuously deformed in TMG for the same reasons as for de Sitter space. The only maximally symmetric spaces that can be infinitesimally homogeneously deformed in TMG are anti-de Sitter, the sphere, and flat spacetime.

Studying the infinitesimal homogeneous deformations of the maximally symmetric spaces quickly reduces to the infinitesimal deformations of all three-dimensional Lie algebras. It is then trivial to extend our original target and talk about all homogeneous, possibly isotropic, deformations of homogeneous anisotropic solutions in TMG. In order to do this, we ask for each anisotropic solution if the full isometry algebra is of dimension larger than three. The result is that the isometry algebra is at most that of the type D solutions, the null warped anti-de Sitter, or the maximally symmetric spaces. Only for the type D solutions and at certain extremal limits of the deformation parameter, we arrive at a non-Einstein spacetime that is homogeneous but not given by a three-dimensional Lie group. These extremal limits give AdS$_2\times \RR$, $S^2\times\RR$ and $H_2\times\RR$, where $H_2$ is the two-dimensional hyperbolic space. They are conformally flat and satisfy the Cotton equation $C_{ab}=0$. In all other cases,
 and in general when not taking such extremal limits, homogeneous deformations of anisotropic homogeneous solutions are again anisotropic homogeneous solutions, where we can either continuously deform the three-dimensional Lie group or deform the metric at the identity of the group. The solution space of anisotropic solutions is quite limited and we can discuss all of these deformations. This can be useful insofar as it gives a picture of continuity to the anisotropic solutions.

Nevertheless, the problem of deformations in any given theory can be quite physical. Assuming we know the physics of an undeformed solution, at $t=0$, we can ask whether similar properties or relations hold when we deform the theory at $t\neq0$. Already there are such proposals in the context of holography. When the Cotton tensor is turned off, by setting $\mu=0$ in \eqref{eq:EinCot}, the unique solution of pure Einstein theory with negative cosmological constant is anti-de Sitter. It is assumed that the theory falls into the usual class of AdS/CFT dualities, see for instance \cite{Carlip:2005zn,Witten:2007kt}. One might expect that by turning on the parameter $\mu$ a holographic duality will continue to hold. The holography at $\mu\neq0$ was the proposal of \cite{Anninos:2008fx} for the spacelike warped anti-de Sitter black holes, as being dual to the states of a yet unknown two-dimensional CFT. The conjecture of a deformed duality lends credence from the matching of the black hole entropy to a version of 
Cardy's universal 2d CFT 
entropy formula. In the asymptotic symmetry analysis of the spacelike warped black holes \cite{Compere:2008cv,Compere:2009zj} only one of the two Virasoro algebra survives in the deformation, along with a current algebra. The properties of a conformal field theory with the requisite symmetries was studied most recently in \cite{Detournay:2012pc}. In \cite{Skenderis:2009nt} it was pointed out that spacelike warped anti-de Sitter cannot be holographically renormalized conventionally, if at all. The holographic renormalization of null warped anti-de Sitter and the higher-dimensional Schr\"odinger backgrounds was investigated earlier in \cite{Guica:2010sw}.

In the context of holography, our problem can be read as follows. By turning on the Cotton tensor explicitly by deforming $\mu$ in \eqref{eq:EinCot}, we find new solutions. If a holographic duality between the bulk theory and a dual CFT survives, we may assume that a vacuum solution has as many exact symmetries as possible. These are the homogeneous solutions, and the anisotropic solutions give a first classification. However, the solutions we find are known and their interpretation in holography have appeared as those of the warped AdS/CFT duality and  Schr\"odinger/CFT duality. The zero cosmological constant holography of \cite{Bagchi:2012xr,Bagchi:2012yk} would be applicable 
to our solutions wherever either $R=0$ identically or $R$ can be set to zero as a free parameter. The latter is then the opposite procedure to a deformation, a contraction. 

The outline of this paper is as follows. In order to identify our solutions, we will describe the landscape of known solutions in section~\ref{sec:known} following \cite{Chow:2009km}, which uses the Segre-Petrov classification into spacetime types. In section \ref{sec:3dLie} we list the classification of three-dimensional Lie algebras that we use. We then solve TMG for each of them in section \ref{sec:3dGeometry}. A summary of our results can be found in the beginning of section \ref{sec:3dGeometry} in table \ref{table:solutions}. In section \ref{sec:LieCoho} we introduce the notion of Lie algebra deformations and describe the three-dimensional Lie algebra deformations. The three-dimensional Lie algebra deformations are concisely summarized in figure~\ref{fig:def3d}. Finally, in section \ref{sec:HomoDef} we explain the homogeneous deformations of the anisotropic solutions. In the discussion of section \ref{sec:discussion} we make some comments on our solutions with respect to the literature. The appendix 
contains some calculations that are needed in section \ref{sec:HomoDef}.

\section{Solutions of TMG}\label{sec:known}
In this section we describe the known solutions of TMG in order to identify the solutions that we find. The solutions to TMG are most easily classified by using the Segre-Petrov type of the traceless Einstein operator 
\begin{equation}
S^a_b=R^a_b-\frac{R}{3}\delta^a_b,
\end{equation}
as was done in~\cite{Chow:2009km}. The Segre type of an operator $S^a_b$ will be denoted $\{a_1\cdots a_n\}$ where the integers $a_i$ are the dimensions of its Jordan blocks. Type $\{1z\bar{z}\}$ means that $S^a_b$ has only one real eigenvalue, type $\{12\}$ that there are two real \emph{geometric} eigenvalues, and so on. When there is a degeneracy of geometric eigenvalues, let us momentarily denote this by surrounding the repective integers $a_i$ with round parantheses and replace $\{\ldots\}$ with $[\ldots]$, e.g. $[1(11)]$ has three geometric eigenvalues and two of these are equal. We will use the Segre-Petrov labeling of~\cite{Chow:2009km}, see table~\ref{table:SP}. For instance type $[1(11)]$ will be called type D. 
 
\begin{table}
\begin{center}
\begin{tabular}[thb]{l*7{c}}
\toprule
Segre-Petrov & O & N & D & III & II & I$_\RR$ & I$_\CC$\\
Segre type   & [(111)] & [(12)] & [(11)1] & [3] & [12] &
[111]& [1$z\bar{z}$]
\\\bottomrule
\end{tabular}
\end{center}
\caption{Three-dimensional Petrov-Segre classification of a traceless operator.} \label{table:SP}
\end{table}

We will use a definition of warped anti-de Sitter that slightly disagrees with that of \cite{Chow:2009km}, but agrees with \cite{Anninos:2008fx}. 
\begin{definition}A left-invariant metric on $\mathrm{SL}(2,\RR)$ will be called \emph{warped anti-de Sitter} if the isometry algebra is enhanced from ${\mathfrak{sl}_2}$ to ${\mathfrak{sl}_2}\oplus\RR$, where $\RR$ is generated by an element $\xi$ of the right action on $\mathrm{SL}(2,\RR)$. It will be called timelike, spacelike or null warped anti-de Sitter when $\xi$ is respectively an elliptic, hyperbolic, or parabolic element of the right action. 
\end{definition}
The metric of spacelike, timelike and null warped anti-de Sitter are respectively
\begin{align}
g_s & = 
\beta\left( - \cosh^2\sigma \,\grad t^2 + \grad\sigma^2 \right)
+\gamma \left( \grad u + \sinh\sigma\,\grad t\right)^2~\label{eq:warpeds}~,\\
g_t & =\gamma\left(+\cosh^2\sigma\, \grad t^2 + \grad\sigma^2\right)
+\alpha \left( \grad u + \sinh\sigma \grad t\right)^2~,\label{eq:warpedt}
\\\intertext{and}
g_n & = 
\alpha\frac{\grad z^2+\grad u \, \grad v}{z^2} \pm \frac{{\grad u}^2}{ z^{4} } ~.\label{eq:warpedn}
\end{align}
The warped anti-de Sitter spaces are respectively of type D, D and N. For positive cosmological constant or for euclidean signature, the metric of the squashed / stretched sphere of either signature also satisfies the equations of TMG. Here and in the rest of the text we will be signature agnostic without specifying signs of parameters such as $\alpha$, $\beta$, $\gamma$ in the metrics. For definiteness, we will restrict to mostly plus signatures\footnote{so for instance, $\gamma$ in \eqref{eq:warpedt} cannot be negative, etc.}, Euclidean or Lorenzian. Similar to the warped anti-de Sitter metrics, we define:
\begin{definition}
 A left-invariant metric on $\mathrm{SU}(2)$ that is furthermore invariant under a one-dimensional subgroup of the right action will be called the \emph{stretched / squashed sphere}.
\end{definition}
In local coordinates the squashed / stretched sphere metric is
\begin{equation}
 g = \alpha\left( \grad \theta^2 + \sin^2\theta \grad\phi^2\right) + \gamma\left(\grad \psi + \cos\theta\grad\phi\right)^2~.
\end{equation}
That is, in our definition it can be of either signature according to the sign of $\gamma$. The squashed / stretched sphere is of type D.

A useful theorem that we will use repeatedly is the one in \cite{Chow:2009km} about type D solutions. We generalize it for arbitrary mostly plus metric signature $\sigma=\pm1$ and for any value of the cosmological constant: if a TMG solution is of type D then
\begin{enumerate}
\item there is a non-null Killing vector $k$ of constant norm $|k|^2=\pm1$ such that 
\begin{equation}\label{eq:typeDdA}
\nabla_a k_b = \frac{\sigma}{3\mu} \epsilon_{abc}k^c~,\end{equation}
\item the Kaluza-Klein reduced two-dimensional metric along $k$ has constant curvature equal to
\begin{equation}\label{eq:typeDR2} R^{(2)} = R +\frac{2}{9\mu^2}\sigma~.\end{equation}
\end{enumerate}
Already, one can use \eqref{eq:typeDdA} and \eqref{eq:typeDR2} to write the type D solutions as
\begin{equation}\label{eq:typeDKK}
 g=g_{(2)} \pm \left(\grad u + A \right)^2~,
\end{equation}
where $g_{(2)}$ has scalar curvature given by \eqref{eq:typeDR2} and $A$ is such that $\grad A = \frac{1}{6\mu}\dvol_{(2)}$. The sign in \eqref{eq:typeDKK} is the sign of the norm of the Killing vector $k=\partial_u$.

For negative cosmological constant, $R<0$, and lorentzian signature, $\sigma=-1$, a type D solution implies negative $R^{(2)}$. One thus obtains the spacelike or timelike warped anti-de Sitter, according to the signature of the two-dimensional base space. However, the theorem can also be satisfied with $R^{(2)}>0$. In this case, the two-dimensional base space is either a two-sphere or two-dimensional de Sitter space. The 3d space over the two-sphere is the squashed / stretched sphere of either signature. The 3d space over two-dimensional de Sitter agrees\footnote{Note that two-dimensional de Sitter is an overall sign change in the metric of two-dimensional anti-de Sitter. Elsewhere, this solution has been called warped de Sitter~\cite{Anninos:2009jt}.} with what we call spacelike warped anti-de Sitter, the metric in \eqref{eq:warpeds} with $\beta<0$. Finally, the theorem can also be satisfied with $R^{(2)}=0$, so that the two-dimensional base space is flat,
\begin{equation}
 g_f = \grad x^2 \pm \,\grad t^2  \pm \left( \grad u + \nu\,x\,\grad t\right)^2~.
\end{equation}
The two signs here are uncorellated and there are three different choices for euclidean or mostly plus lorentzian signature. Following~\cite{Anninos:2009jt}, we call the latter  metric as that of \emph{warped flat}. The type D solutions are thus warped spacelike and timelike anti-de Sitter, the squashed / stretched sphere and warped flat space. 

Setting the general Kundt solutions aside, the ubiquitous solutions with negative cosmological constant are the warped anti-de Sitter solutions and the anti-de Sitter generalization of a pp-wave. In order to identify some of the solutions we find, we will need the form of the general TMG pp-wave metric in anti-de Sitter. For ${\mu^2}\neq-6/R$ the pp-wave is written in~\cite{Gibbons:2008vi} as
\begin{equation}\label{eq:oldpp} g =\ell^2\left( \grad \rho^2 +2\, e^{2 \rho} \, \grad u \, \grad v \right) + e^{(1\pm\frac{\ell}{\mu})\rho} f(u) \grad u^2 ,\end{equation}
where $\ell=\sqrt{-6/R}$ is the anti-de Sitter radius and both signs in the exponent give solutions for an appropriate sign of the volume form. For $\mu^2=\ell^2$ there are
 again  two solutions:
\begin{align}
g & =  \grad \rho^2 + 2\,e^{2\rho}\grad u \grad v + \rho\, e^{2\rho}f(u)\,\grad u^2  ~.\label{eq:ppcrit1}  \\\intertext{and}
g & =  \grad \rho^2 + 2\,e^{2\rho}\grad u \grad v + \rho f(u)\grad u^2~.
\end{align}
The general pp-wave solution is of type N. In our terminology and following~\cite{Anninos:2010pm}, the metric \eqref{eq:oldpp} at $\ell=3\mu$, with a constant $f(u)=\pm1$ and with the positive sign in the exponent of the component $g_{uu}$, corresponds to what we call the null warped anti-de Sitter. 

When the cosmological constant vanishes, the homogeneous anisotropic solutions are given in the works of \cite{Ortiz:1989vc} and \cite{Nutku:1989qi}. There are a few more solutions that are not the special $R=0$ cases of warped anti-de Sitter, the pp-waves, or the squashed / stretched sphere. These solutions are of type I$_\RR$, which are sometimes called triaxially deformed anti-de Sitter or triaxially deformed sphere, type II and type I$_\CC$. In our work we take a generically non-vanishing scalar curvature $R$. Therefore, we do recover these solutions for when $R=0$. They all appear as left-invariant metrics on $\mathrm{SL}(2,\RR)$. 

Finally, there are the Kundt solutions of TMG that were solved for in~\cite{Chow:2009vt}. In three dimensions, the Kundt condition reduces to requiring a null expansion-free  geodesic congruence. The general solution is quite involved. It is relevant here to describe only the constant scalar invariant (CSI) Kundt solutions. A CSI spacetime is such that all curvature invariants are constant. There are two families of CSI Kundt solutions that are generically of type II and III. The family of type II can reduce to type D. The family of type III breaks up into a number of subfamilies, in particular three subfamilies when the cosmological constant is negative. Each subfamily can reduce to type N, and one of the type N subfamilies contains the pp-wave. 

\section{Three-dimensional Lie algebras}\label{sec:3dLie}
We will present our own labeling scheme for the three-dimensional Lie algebras. We include the abelian $\RR^3$ and the two simple algebras ${\mathfrak{sl}_2}$ and $\mathfrak{su}_2$. We also have the Lie algebras $\mathfrak{a}_0$ and $\mathfrak{a}_\infty$, and two continuous families of Lie algebras: $\mathfrak{\mysosmall}(1,1;\theta)$ and $\mathfrak{\mysosmall}(2;\theta)$. The parameter $\theta$ takes values in $[0,\frac{\pi}{2}]$ and $\mathfrak{\mysosmall}(1,1;0)=\mathfrak{\mysosmall}(2;0)$.  We relate these algebras to the Bianchi classification~\cite{Bianchi,ModernBianchi,EllisMacCallum} in table~\ref{table:bianchisimple}, see the appendix \ref{app:groupdiff} for some other conventions.  

\begin{table}[thb]
\begin{center}
\begin{tabular}[h]{*{2}{l}}
\toprule
Bianchi       & here\\\midrule
I &  $\RR^3$  \\
II&    $\mathfrak{a}_\infty$ \\
III (VI$_{-1}$) & $\mathfrak{\mysosmall}(1,1;\frac\pi4) $\\
IV &  $\mathfrak{a}_0$  \\
V&  $\mathfrak{\mysosmall}(2;0)$\\
VI$_{h\leq 0}$  &  $\mathfrak{\mysosmall}(1,1;\theta),\,\theta\in(0,\frac\pi2]$\\
VII$_{h\geq 0}$    & $\mathfrak{\mysosmall}(2;\theta),\,\theta\in(0,\frac\pi2]$ \\
VIII&   ${\mathfrak{sl}_2}$  \\
IX  & $\mathfrak{su}_2$   \\
\bottomrule
\end{tabular}
\end{center}
\caption{Classification of three-dimensional Lie algebras. The Behr invariant $h$ is as in \protect\cite{EllisMacCallum}, with 
$h=\pm\cot^2\theta$ for respectively $\mathfrak{\mysosmall}(2;\theta)$ and $\mathfrak{\mysosmall}(1,1;\theta)$.} 
\label{table:bianchisimple}
\end{table}

Let us first describe the continuous families. The Lie algebra $\mathfrak{\mysosmall}(2;\theta)$  is spanned by $l$, $m_1$ and $m_2$, with the only non-vanishing brackets being
\begin{align}\label{eq:commso2}
 [l,m_1]&= 2\cos\theta\, m_1 + 2\sin\theta\, m_2 ~,&
 [l,m_2]&= 2\cos\theta\, m_2 - 2\sin\theta\, m_1 .
\end{align}
That is, $l$ acts as a euclidean rotation twisted by a simultaneous dilation on the vector $\{m_i\}\in\RR^2$. The second family $\mathfrak{\mysosmall}(1,1;\theta)$ is spanned by $l'$, $m'_1$ and $m'_2$, with non-vanishing brackets
\begin{align}\label{eq:commso11}
 [ l',m'_1]&= 2\cos\theta\, m'_1 + 2\sin\theta\, m'_2~, &
 [ l',m'_2]&= 2\cos\theta\, m'_2 + 2\sin\theta\, m'_1 .
\end{align}
That is, $l'$ acts as a lorentzian rotation twisted by a dilation on the vector $\{m'_i\}\in\RR^{1,1}$. Both Lie algebras limit at $\theta=0$ to the same Lie algebra $\mathfrak{\mysosmall}(1,1;0)=\mathfrak{\mysosmall}(2;0)$, in which the action of $l$ on $m_1$ and $m_2$ is a rescaling by the {same factor}. 
We will drop for simplicity the primes in \eqref{eq:commso11}.

In order to define $\mathfrak{a}_0$ and $\mathfrak{a}_\infty$, it is convenient to introduce a continuous family of Lie algebras $\mathfrak{a}_\lambda$, $\lambda\in\RR$. Let it be spanned by $r$, $x$ and $y$, and with non-vanishing brackets
 \begin{align*}
 [r,x]&=x-\lambda\,y~,&[r,y]&=x+y~. 
 \end{align*}
The Lie algebra $\mathfrak{a}_0$ is precisely given by $\mathfrak{a}_\lambda$ at the value of $\lambda=0$. The Lie algebra $\mathfrak{a}_\infty$ can be thought of as a limit of the family: first rescale the basis by $(r,x,y)\mapsto \frac1\lambda\,(r,x,y)$ and then send $\lambda$ to infinity. The only non-vansishing Lie bracket of the limit $\mathfrak{a}_\infty$ is $[r,x]=-y$.

\begin{figure}
\begin{center}
\includegraphics{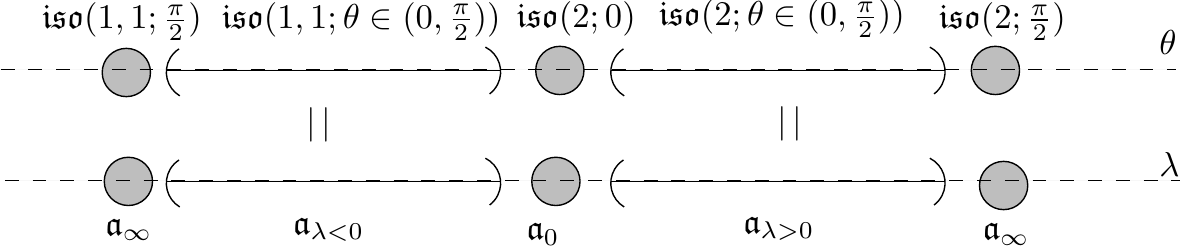}
\caption{Comparing the continuous families $\mathfrak{a}_\lambda$ with $\mathfrak{\mysosmall}(1,1;\theta)$ and $\mathfrak{\mysosmall}(2;\theta)$.}\label{fig:figaso}
\end{center}
\end{figure}

One can show that the family $\mathfrak{a}_\lambda$ for $\lambda<0$ is isomorphic to the family $\mathfrak{\mysosmall}(1,1;\theta)$ with $\lambda=-\tan^2\theta$ and the family $\mathfrak{a}_\lambda$ with $\lambda>0$ is isomorphic to $\mathfrak{\mysosmall}(2;\theta)$ with $\lambda=\tan^2\theta$. However $\mathfrak{a}_0$ is not isomorphic to $\mathfrak{\mysosmall}(1,1;0)=\mathfrak{\mysosmall}(2;0)$. Nor is any of the Lie algebras $\mathfrak{a}_\infty$, $\mathfrak{\mysosmall}(2;\frac{\pi}{2})$ and $\mathfrak{\mysosmall}(1,1;\frac{\pi}{2})$ isomorphic to each other. This is presented pictorially in figure \ref{fig:figaso}.  Any three-dimensional Lie algebra is isomorphic to one of 
$\RR^3$, 
${\mathfrak{sl}_2}$, 
$\mathfrak{su}_2$, 
$\mathfrak{a}_0$, 
$\mathfrak{a}_\infty$, 
$\mathfrak{\mysosmall}(1,1;\theta)$, 
$\theta\in(0,\frac\pi2]$, and $\mathfrak{\mysosmall}(2;\theta)$, $\theta\in[0,\frac\pi2]$. An alternative classification of three-dimensional Lie algebras is thus given by $\mathfrak{a}_\lambda$,
$\lambda\in\RR$,
$\mathfrak{a}_\infty$,
$\mathfrak{\mysosmall}(1,1;\frac{\pi}{2})$, 
$\mathfrak{\mysosmall}(2;\frac{\pi}{2})$, $\mathfrak{\mysosmall}(2;0)$, and of course $\RR^3$,  ${\mathfrak{sl}_2}$ and $\mathfrak{su}_2$. Note also the isomorphisms $\mathfrak{\mysosmall}(1,1;\frac{\pi}{2})\approx \mathfrak{so}(1,1)\ltimes \RR^{1,1}$,  $\mathfrak{\mysosmall}(2;\frac{\pi}{2})\approx\mathfrak{so}(2)\ltimes \RR^2$ and $\mathfrak{\mysosmall}(1,1;\frac{\pi}{4})\approx\RR\oplus\left(\RR\ltimes\RR\right)$.

\section{Geometry of three-dimensional groups}
\label{sec:3dGeometry}
In this section, for each three-dimensional Lie algebra $\mathfrak{g}$ of a group $G$, we fix a basis $\tau_a$ in which the structure coefficients are given by $[\tau_a,\tau_b]=f_{ab}{}^c\tau_c$. The basis $\tau_a$ induces the left-invariant vector field basis $r_a$ that generate the right action, with $[r_a,r_b]=-f_{ab}{}^c r_c$, the right-invariant vector field basis $l_a$ that generate the left action, with $[l_a,l_b]=f_{ab}{}^c l_c$, and the left-invariant Maurer-Cartan one-form basis $\theta^a$ that are dual to the $r_a$, $\theta^a (r_b)=\delta^a_b$. A left-invariant metric $g$ on the group $G$ is given by a non-degenerate metric $B$ on $\mathfrak{g}$,
\begin{equation}
g=B(\theta,\theta)=B_{ab}\theta^a\theta^b~.
\end{equation}
However, the metric $g$ is unique up to the action of the automorphism group on $B$. By using the action of the automorphism group we fix the metric $B$ into classes where in each class the metric $B$ depends on a small set of continuous parameters $(\alpha,\beta,\ldots)$.

We will use the basis of left-invariant one-forms $\theta^a$ and all (pseudo)-Riemannian calculations become algebraic. The metric compatibility of the Levi-Civita connection, with $\nabla_{r_a} \theta^b = - \Gamma^b_{ac}\theta^c$, is solved by
\begin{equation} B_{dc} \Gamma^d_{ab} = B_{db} \Gamma^d_{[ca]} + B_{da} \Gamma^d_{[cb]} + B_{dc} \Gamma^d_{[ab]}~, \end{equation}
where $\Gamma^a_{[bc]} =- \frac{1}{2} f_{bc}{}^a$ is given in terms of the structure coefficients. 
The Riemann curvature 
\[ R_{ab}{}^c{}_d \,\theta^d = -\nabla_{r_a}\nabla_{r_b}  \theta^c +
\nabla_{r_b}\nabla_{r_a}  \theta^c +
\nabla_{[r_a,r_b]} \theta^c\] is
given by
\begin{equation}
  R_{ab}{}^c{}_d = -\Gamma^e_{ad} \Gamma^c_{be} + \Gamma^e_{bd} \Gamma^c_{ae} + f_{ab}{}^e  \Gamma_{ed}^c.
\end{equation}
Indices will be raised and lowered with $B_{ab}$. The Levi-Civita coefficients $\Gamma_{ab}^c$ are thus equivalently given by $\nabla_{r_a}r_b = \Gamma^c_{ab} r_c$.

We test when the metric $B$ satisfies the Einstein-Cotton equation:
\begin{equation}\label{eq:EinCotton} R_{ab} - \frac{R}{3} B_{ab} = \mu \,\epsilon_{a}{}^{cd}\nabla_c R_{bd}~ ~.\end{equation}
We do this by using computer algebra. First we replace $R$ into the equation as a function of the parameters in $B$. We then treat all parameters $(\alpha,\beta,\ldots)$ in $B$, the parameter $\theta$ if the Lie algebra is parametrized by it, and $\mu$ as unknown. Note that the equations are linear in $\mu$. We could have fixed $\mu$ or $R$ up to scale, by using a homothety, but instead we give them as functions of the parameters in $B$ when such a solution exists. The non-Einstein solutions are shown in table~\ref{table:solutions}

\begin{table}[thb]
\centering 
\begin{tabular}{llll}
\toprule
group&solution&characterization& type
\\
\midrule
 {${\mathfrak{sl}_2}$}&
\ref{item:sl2warped} & warped s,t,n 
& D,D,N\\
& \ref{item:sl2triaxial}& $R=0$ triaxially deformed AdS
& I$_\RR$\\
&\ref{item:sl2CSIKundt} & $R=0$ CSI Kundt type II
& II \\
&\ref{item:sl2typeIC} & $R=0$ type $\{1z\bar{z}\}$ & I$_\CC$
\\
 {$\mathfrak{su}_2$}&
\ref{item:su2warped} & squashed / stretched $S^3$ & D \\
& \ref{item:su2triaxial}& triaxially deformed $S^3$
& I$_\RR$
\\
{$\mathfrak{\mysosmall}(1,1;\theta)$}
&
\ref{item:so11warpedA} & ($\theta=\frac\pi4$) warped s,t 
&D \\
&
 \ref{item:so11warpedB} & ($\theta=\frac\pi4$) warped flat 
&D \\
& \ref{item:so11ppA}, \ref{item:so11ppB} & ($\theta\neq0$) pp-waves
&N
\\
{$\mathfrak{a}_\infty$ }
&
\ref{item:flatwarped} & warped flat
&D
\\
{$\mathfrak{a}_0$ }
&
\ref{item:a0pp} & $|\mu|=|\ell|$ pp-wave
& N
\\\bottomrule
\end{tabular}
\caption{All homogeneous anisotropic $\mu\neq0$ TMG solutions.}
\label{table:solutions}
\end{table}

We also test for when a left-invariant metric $B$ on a three-dimensional group $G$ with Lie algebra $\mathfrak{g}$ has more isometries than the left action. At the identity, an extra Killing vector $\xi$ can be made such that it is zero, $\xi|_e=0$, but its first derivative $\nabla\xi|_e$, the linear isotropy, is not zero. At the identity then, the endomorphism $\nabla\xi|_e:\mathfrak{g}\rightarrow\mathfrak{g}$ should leave invariant the metric and all curvature tensors. 
Let $d\in\mathfrak{g}$ be a vector that is left invariant by $\nabla\xi|_e\in\mathfrak{so}(B)$. A necessary condition for the existence of more isometries is that 
there is a left-invariant one-form $d_a$ such that
\begin{equation}\label{eq:killingtest}
 R_{ab} = c_1\, B_{ab} +c_2\, d_a d_b .
  \end{equation}
If $c_2=0$ then the space is maximally symmetric, so assume \eqref{eq:killingtest} is satisfied with $c_2\neq0$. This is equivalent to the spacetime type being D or N, respectively for $B(d,d)\neq0$ or $B(d,d)=0$. It also follows that if the isometry of a left-invariant metric on $G$ is enhanced and is not maximally symmetric, then the isometry is four-dimensional. Indeed, if $B$ is lorentzian, it is easy to show that there is no symmetric bilinear that is invariant under the two-dimensional non-abelian subgroup of $\mathrm{SO}(1,2)$ alone. This is also true for euclidean signature, whereby the only subgroup of $\mathrm{SO}(3)$ up to conjugacy is one-dimensional.

The condition \eqref{eq:killingtest} is sufficient for type D solutions, in which case the type-D theorem of \cite{Chow:2009km} determines the solutions in terms of $R$, $\mu$ and signs in the signature of the metric. The condition \eqref{eq:killingtest} is certainly weak for type N spacetimes. Since $\nabla\xi|_e$ leaves invariant $B_{ab}$, $R_{ab}$ and all higher derivative curvature tensors, then it is also necessary that
\begin{equation}\label{eq:killingtest2}
 \nabla_a d_b = c_3\, \sqrt{B}\, \epsilon_{ab}{}^c\, d_c + c_4 \,B_{ab} + c_5\, d_a d_b~.
\end{equation}
There are no more conditions on the homogeneity of curvature invariants. 

\subsection{Metrics on \texorpdfstring{$\mathrm{SL}(2,\RR)$}{SL(2,R)}}\label{sec:sl2}
Let us fix a basis $\{\tau_0,\tau_1,\tau_2\}$ for the Lie algebra ${\mathfrak{sl}_2}$ with
\begin{align}\label{eq:sl2comm}
 [\tau_0,\tau_1]&=\tau_2 ~,&
 [\tau_2,\tau_1]&=\tau_0~,&
 [\tau_2,\tau_0]&=\tau_1~.
\end{align}
In lorentzian indices, \eqref{eq:sl2comm} is $[\tau_a,\tau_b]=\epsilon_{ab}{}^c\tau_c$ with the antisymmetric symbol $\epsilon_{012}=1$ and the indices in this relation are raised or lowered with a mostly plus metric $\eta_{ab}$. The basis $\tau_a$ induces the left-invariant vector fields $r_a$ and the left-invariant Maurer-Cartan one-forms $\theta^a$. A left-invariant metric $g$ on $\mathrm{SL}(2,\RR)$ is given by a non-degenerate metric $B$ on ${\mathfrak{sl}_2}$, $g=B_{ab}\theta^a\theta^b$, and is unique up to the action of $B\mapsto S^t B S$ with $S\in \mathrm{SO}(1,2)$. The bi-invariant metric of anti-de Sitter is given by the Killing form $B_{ab}=\eta_{ab}$.

Under the action of the automorphism group on $B$ we arrive at four classes that can be labeled by the Segre type of $\eta^{ac}B_{cb}$. They are
\begin{enumerate}
\item Type $\{111\}$.  The metric is given by
\begin{equation}\label{eq:segre111}
g=\alpha \, \theta^0\,\theta^0+\beta\,\theta^1\,\theta^1+\gamma\,\theta^2\,\theta^2 
\end{equation}
and can be of any signature depending on the sign of $\alpha\beta\gamma\neq 0$.
\item Type $\{12\} $. The metric is
\begin{equation}\label{eq:segre12}
g=\beta\left(-\theta^0\,\theta^0+\theta^1\,\theta^1\right)+ \gamma \,\theta^2\,\theta^2 +\delta \left(\theta^0+\theta^1\right)^2 \text{ with }\beta\gamma\delta\neq0~, 
\end{equation}
and it is always lorentzian. It is mostly plus for $\gamma>0$. Here, $\delta$ can be rescaled freely and the metric depends only on its sign. It is a $\delta$-deformation of type $\{111\}$ with $\alpha=-\beta$.
\item Type $\{3\}$. The metric is given by the matrix
\begin{equation} g = \begin{pmatrix} 0&\alpha&\delta\\\alpha&0&0\\\delta&0&2\alpha \end{pmatrix} \text{ with } \alpha\delta\neq0, \end{equation}
in the basis $\{ \theta^0+\theta^1 , -\theta^0+\theta^1,\theta^2 \}$. One can use Descartes' rule of signs\footnote{The rule says that the number of positive roots of the polynomial is either equal to the number of sign differences between consecutive nonzero coefficients of the monomials in decreasing degree of $\lambda$, or is less than it by a multiple of 2. We apply the rule to both $\det(B-\lambda)=0$ and $\det(B+\lambda)=0$.} to show that the signature is always lorentzian. It is mostly plus for $\alpha>0$. It is a $\delta$-deformation of type $\{111\}$ with $-\alpha=\beta=\gamma$ (anti-de Sitter).
\item Type $\{1z\bar{z}\}$. 
\begin{equation}\label{eq:segre1zz}
g=\alpha\left(-\theta^0\,\theta^0+\theta^1\,\theta^1\right)+2\delta\,\theta^0\,\theta^1+\gamma\,\theta^2\,\theta^2\text{ with }
 \alpha\delta\gamma\neq0~.
\end{equation}
The metric is always lorentzian and mostly plus for $\gamma>0$. It is a $\delta$-deformation of type $\{111\}$ with $\alpha=-\beta$.
\end{enumerate}
The classfication here is essentially an adaptation of the one for three-dimensional energy-momentum tensors~\cite{Hall:3dSegre}. A difference to \cite{Hall:3dSegre} is that there the action of $\mathrm{SO}(1,2)$ comes from the orthonormal frame structure, here the action is of the group automorphisms.

\subsubsection*{Equations of Motion}
We test when the metrics \eqref{eq:segre111}-\eqref{eq:segre1zz} for each Segre type satisfy the Einstein-Cotton equation. They are
\begin{ECsols}
\item The so-called warped solutions: \label{item:sl2warped}
\begin{ECsols}
\item The metric of type $\{111\}$ and $\beta=\gamma$, called 
timelike warped AdS, in which case
\begin{align}
R&=-\frac{\alpha+4\gamma}{2\gamma^2}~,&
\mu^2 &= \frac{4\gamma^2}{9|\alpha|},
\end{align}
\item The metric of type $\{111\}$ and $\alpha=-\beta$, called 
spacelike warped  AdS, in which case
\begin{align}
R &= \frac{-4\beta+\gamma}{2\beta^2}~,
&
\mu^2  &= \frac{4\beta^2}{9|\gamma|}~,\label{eq:RmuSWAdS}
\end{align}
\item The metric of type $\{12\}$ and $\beta=\gamma$, called null warped AdS, in which case ($\gamma>0$)
\begin{align}
R&=-\frac{3}{2\gamma}~, &
\mu^2\,R&=-\frac{2}{3}~,
\end{align}
\end{ECsols}
\item\label{item:sl2triaxial} 
The metric of type $\{111\}$ whenever $(\alpha+\beta+\gamma)^2=4\beta\gamma$. In this case $R=0$ and
\begin{equation} \mu^2 = \frac{{|\alpha\beta\gamma|}}{(\alpha-\beta-\gamma)^2} ~.\end{equation}
\item\label{item:sl2CSIKundt} The metric of type $\{12\}$ and 
 $\gamma=4\beta$, in which case $R=0$ and $\mu^2=|\beta|/9$.
\item\label{item:sl2typeIC} The metric of type $\{1z\bar{z}\}$ whenever $\left(\frac\gamma2-\alpha\right)^2=\alpha^2+\delta^2$, in which case $R=0$ and
\begin{equation}
\mu^2=\frac{{|\gamma|(\alpha^2+\delta^2)}}{(\gamma+2\alpha)^2}
~.
\end{equation}
\end{ECsols}
In each of these solutions, the parameter $\mu$ is determined algebraically by the parameters in $B$. Converseley, any choice of $\mu\in\RR$ gives any of the above solutions due to the scaling properties of the Cotton tensor: $g_{ab}\mapsto \Lambda^2 g_{ab}$, $C_{ab}\mapsto \Lambda^{-1}C_{ab}$ and $\mu\mapsto\Lambda\mu$. 

We see that the generically $R\neq0$ solutions are the known spacelike, timelike and null warped anti-de Sitter. 
The TMG solutions on $\mathrm{SL}(2,\RR)$ beyond the ``warped'' ones, cases \ref{item:sl2triaxial}, \ref{item:sl2CSIKundt} and \ref{item:sl2typeIC}, satisfy the Ricci-Cotton equation $R_{ab}=\mu\,C_{ab}$ with $R=0$. Therefore, they should be included in the work of \cite{Ortiz:1989vc}. For instance, \ref{item:sl2triaxial} can be rewritten so that it manifestly matches Ortiz's solution (a) (case $a=0$). We will also show that \ref{item:sl2CSIKundt} corresponds to a CSI Kundt solution in the family of spacetime type II. On the other hand, the solutions \ref{item:sl2triaxial} and \ref{item:sl2typeIC} are not Kundt: \ref{item:sl2triaxial} is generically of type $I_\RR$ and \ref{item:sl2typeIC} is always\footnote{More generally, all metrics of type $\{1z\bar{z}\}$ are of type I$_\CC$.} of type $I_\CC$. The solutions  \ref{item:sl2triaxial} and \ref{item:sl2typeIC} have the same spacetime type as Ortiz's (b) solution. We will therefore identify them, see also section \ref{sec:discussion}. If one fixes $R=0$,
 then the spacelike or timelike warped anti-de Sitter solutions match the \ref{item:sl2triaxial} solution at two separate solutions found by Vuorio~\cite{Vuorio:1985ta}.

\subsubsection*{Identifying \ref{item:sl2CSIKundt}}
The Ricci tensor $R^a_b$ of \ref{item:sl2CSIKundt} is of type II. Other than in \cite{Ortiz:1989vc}, type II CSI metrics have only reappeared as the CSI Kundt of type II~\cite{Chow:2009vt}, with metric
\begin{equation}\label{eq:csikundtsl}
g = 2 \grad u \grad v - \frac{1}{9}\left(\frac{1}{\mu^2}-\frac{R}{9}\right) v^2 \grad u^2
+ \left(d\rho+\frac{2}{3}\frac{v}{\mu}\grad u\right)^2 + f_0(u,\rho)\grad u^2 ~,
\end{equation}
and $f_0(u,\rho)$ satisfies the differential equation
\begin{equation}\label{eq:finKII}
\partial_\rho^3 f_0+\frac{2}{\mu}\partial_\rho^2 f_0+(\frac{5}{9\mu^2}+\frac{R}{2})\partial_\rho f_0=0~.
\end{equation}
One can gauge away the constant in $\rho$ term of $f_0$. In our case we also set $R=0$.

We use the (locally defined) ``extremal'' coordinates of $\mathrm{SL}(2,\RR)$, see  \cite{Jugeau:2010nq}, whereby elements of $\mathrm{SL}(2\,\RR)$ are parametrized by 
\begin{equation}
\mathcal{V}(x)=e^{t(\tau_0+\tau_2)}e^{\sigma\tau_1}e^{z \tau_2}~.
\end{equation}
The Maurer-Cartan one-forms are 
\begin{equation}\label{eq:accelV}
\mathcal{V}^{-1}\grad \mathcal{V} =
( e^{\sigma}\cosh z\, \grad t - \sinh z \,\grad \sigma)\,\tau_0
+ (\cosh z \,\grad \sigma-  e^\sigma\sinh z \,\grad t)\,\tau_1+ (\grad z+e^{\sigma}\grad t)\,\tau_2~.
\end{equation}
In these coordinates, the spacelike warped AdS metric with $-\alpha=\beta=\gamma/4$ is given by
\begin{equation}\label{eq:accelswbase}\beta\left( -e^{2\sigma} \grad t^2 + \grad \sigma^2 + 4\left(\grad z + e^\sigma \grad t\right)^2\right)~.\end{equation}
The solution  \ref{item:sl2CSIKundt} is a deformation of this by the addition of the following term:
\begin{equation}\label{eq:accelswdef}
\delta\, (\theta^0 + \theta^1)^2 =\delta\, e^{2 z}(\grad \sigma +e^{\sigma} \grad t )^2~.\end{equation}
The change of coordinates from \eqref{eq:accelswbase} and \eqref{eq:accelswdef} to those in \eqref{eq:csikundtsl} is given by
\begin{align}
t&=\frac{1}{2 v} + \frac{1}{18}(\frac{1}{\mu^2}-\frac{9R}{2}) u \\
e^\sigma &= 2 v\\
z &= \frac{1}{6\mu}\left(\frac{1}{\mu^2}-\frac{9R}{2}\right)\rho + \ln v ~.
\end{align}
This way we confirm that $f_0(u,\rho)=c_1\,e^{-\rho/3\mu}$, which indeed satisfies \eqref{eq:finKII}.

\subsubsection*{Extra isometries}
The tests \eqref{eq:killingtest} and \eqref{eq:killingtest2} on the $\mathrm{SL}(2,\RR)$-invariant metrics reproduce only the warped AdS spaces. An extra Killing vector is given by the commuting left-invariant vector: $r_2$ for spacelike warped, $r_0$ for timelike, or $r_0+r_1$ for null warped. 
We will show by example that among all Einstein-Cotton Lie group solutions, the isometry is enhanced to a four-dimensional algebra only for the warped solutions, the squashed / stretched sphere, and their flat base limit. 

\subsection{Metrics on \texorpdfstring{$\mathrm{SU}(2)$}{SU(2)}}
Let us fix a basis $\{\tau_1,\tau_2,\tau_3\}$ for the Lie algebra $\mathfrak{su}_2$ with $[\tau_i,\tau_j]=\epsilon_{ij}{}^{k}\tau_k$. The structure coefficients are given by $\epsilon_{123}=1$ and indices in this relation are raised and lowered with the euclidean metric $\delta_{ij}$. The basis $\tau_i$ induces the bases of the left-invariant vector fields $r_i$ and their dual left-invariant Maurer-Cartan one-forms $\theta^i$, $\theta^i(r_j)=\delta^i_j$. A left-invariant metric $g$ on $\mathrm{SU}(2)$ is given by a non-degenerate metric $B$ on $\mathfrak{su}_2$,
\begin{equation}
g=B(\theta,\theta)=B_{ij}\theta^i\theta^j~,
\end{equation}
up to the action of the automorphism group $\mathrm{SO}(3)$. Since symmetric matrices are diagonalizable by $\mathrm{SO}(3)$, we take the metric to be
\begin{equation}
g=\alpha \, \theta^1\,\theta^1+\beta\,\theta^2\,\theta^2+\gamma\, \theta^3\,\theta^3~.
\end{equation}
We assume a euclidean or mostly plus lorentzian metric, depending on the sign of $\alpha\beta\gamma\neq 0$.

The metric $g$ is Einstein for $\alpha=\beta=\gamma$. It is the metric of the sphere with isometry algebra $\mathfrak{su}_2\oplus\mathfrak{su}_2$. The Einstein-Cotton equation is solved, besides the Einstein solution, if and only if
\begin{ECsols}
\item \label{item:su2warped} two of the parameters $(\alpha, \beta, \gamma)$ are equal, for definiteness say $\alpha=\beta$. This is the so-called stretched / squashed sphere deformed over its Hopf fibration. We have 
\begin{align}
R&=\frac{\gamma-4\beta}{2\beta^2}~,
&
\mu^2&=\frac{4}{9}\frac{\beta^2}{|\gamma|}~.
\end{align}
The signature of the metric depends on the sign of $\gamma$.
\item \label{item:su2triaxial} whenever $(\alpha+\beta-\gamma)^2=4\alpha\beta$. The metric is always euclidean, with $R=0$ and 
\begin{equation}
\mu^2=\frac{|\alpha\beta\gamma|}{(\alpha+\beta+\gamma)^2}~.
\end{equation}
\end{ECsols}
The isometry algebra of a left-invariant metric on $\mathrm{SU}(2)$ enhances only for the stretched / squashed sphere, in which case it is either $\mathfrak{su}_2\oplus\RR$ or bi-invariant for the round sphere. The stretched / squashed sphere is of type D. We note that the two solutions, \ref{item:su2warped} and \ref{item:su2triaxial} match for $\alpha=\beta=\gamma/4$ at the Vuorio solution~\cite{Vuorio:1985ta}. When not such, the solution\footnote{In general, the triaxially deformed metric can still be of type D for some other values, but will not be a TMG solution.} \ref{item:su2triaxial} is of type I$_\RR$.

\subsection{Metrics on \texorpdfstring{$\mathrm{\mysocapital}(1,1;\theta)$}{\mysocapital(11;theta)}}\label{sec:so11}
Recall the basis $\{l,m_1,m_2\}$ of $\mathfrak{\mysosmall}(1,1;\theta)$,
\begin{align*}
 [ l,m_1]&= 2\cos\theta\, m_1 + 2\sin\theta\, m_2~, &
 [ l,m_2]&= 2\cos\theta\, m_2 + 2\sin\theta\, m_1 ,
\end{align*}
where we have dropped the primes in \eqref{eq:commso11}. We label the dual basis by $\{\tilde{l},\tilde{m}_1,\tilde{m}_2\}$. A left-invariant metric on $\mathrm{\mysocapital}(1,1;\theta)$ is given by a metric $B$ on $\mathfrak{\mysosmall}(1,1;\theta)$ up to the automorphism group. For all values of $\theta\neq0,\frac\pi2$, we have
\begin{equation}\label{eq:autso11}
\mathrm{Aut}(\mathfrak{\mysosmall}(1,1;\theta))=
\left(\ZZ_2\times\mathrm{SO}(1,1)\times\RR^+\right)\ltimes \RR^{1,1}~.
\end{equation}
The $\ZZ_2$ in \eqref{eq:autso11} flips $(m_1,m_2)\mapsto(m_2,m_1)$. The $\mathrm{SO}(1,1)$ and dilation $\RR^+$ act on the $\{m_i\}\in\RR^{1,1}$ in the fundamental representation of $\mathrm{SO}(1,1)$. Finally, the $\RR^{1,1}$ automorphisms send $l\mapsto l+\sum_{i=1}^2x_i\, m_i$. When $\theta=\frac\pi2$ the $\mathrm{SO}(1,1)$ automorphisms are enhanced to $\mathrm{O}(1,1)$ where we can send $(l,m_1,m_2)\mapsto(-l,m_1,-m_2)$. We will use the automorphisms to fix the most general metric. However, we need to consider two cases, which lead to the two metrics $B_1$ and $B_2$.

The $\RR^{1,1}$ automorphisms act on the dual space as $\tilde{m}_i\mapsto \tilde{m_i}-x_i\,\tilde{l}$. If the matrix $B(m_i,m_j)$ is non-degenerate, we can use these to arrange for $B(l,m_1)=B(l,m_2)=0$. The metric thus far is given by\footnote{\label{page:footZ} 
The $\ZZ_4$: $(m_1,m_2)\mapsto (-m_2,m_1)$ and $\ZZ_2: (m_1,m_2)\mapsto(m_1,-m_2)$ send $\theta\mapsto -\theta$, which along with $l\mapsto -l$ is used to fix $0\leq\theta\leq\frac\pi2$. This means that in the metric \eqref{eq:so11metric}, we can only exchange $\alpha$ with $\beta$ when $\theta=\frac\pi2$.}
\begin{equation}\label{eq:so11metric}
 B_{1} = \delta \,\tilde{l}\,\tilde{l} + \alpha \, \left(\tilde{m}_1+\tilde{m}_2\right)^2 +\beta \,\left(\tilde{m}_1-\tilde{m}_2\right)^2+ 2\gamma\,\left(\tilde{m}_1\,\tilde{m}_1-\tilde{m}_2\,\tilde{m}_2\right) ~,
 \end{equation}
with $\gamma^2\neq \alpha\beta$. The $\mathrm{SO}(1,1)$ automorphism rescales $(\alpha,\beta)\mapsto (\Lambda^2 \alpha,\Lambda^{-2}\beta)$ and $\RR^+$ rescales $(\alpha,\beta,\gamma)\mapsto \Lambda^2(\alpha,\beta,\gamma)$. We may therefore fix at least two parameters of $(\alpha,\beta,\gamma)$ that appear in the metric, if non-zero, up to their sign to be $\pm1$. We are signature agnostic and take $\delta\neq0$. However, if $\gamma^2-\alpha\beta>0$, then the metric is lorentzian and the mostly plus signature is for $\delta>0$.

If the matrix $B(m_i,m_j)$ is degenerate, then we can no longer bring the metric to the form $B_1$. However, we can still use the $\ZZ_2\times\ZZ_2\times\RR^+\ltimes\RR^{1,1}$ automorphisms to arrange for $B(l,m_2)=0$ and $B(l,m_1)=1$. The first $\ZZ_2$ is the one in \eqref{eq:autso11} and the second one is the parity transformation in $\mathrm{SO}(1,1)$. The metric thus far is given by
\begin{equation}\label{eq:so11metricsemi}
 B_{2} = \delta \,\tilde{l}\,\tilde{l} + \tilde{l}\,\tilde{m}_1+ \alpha \, \left(\tilde{m}_1+\tilde{m}_2\right)^2 +\beta \,\left(\tilde{m}_1-\tilde{m}_2\right)^2+ 2\gamma\,\left(\tilde{m}_1\,\tilde{m}_1-\tilde{m}_2\,\tilde{m}_2\right) ~,
 \end{equation}
with $\gamma^2 = \alpha\beta$ and $\alpha+\beta\neq2\gamma$ so that $B_2$ is non-degenerate. We can then use the $\mathrm{SO}(1,1)$ automorphisms to rescale $(\alpha,\beta)\mapsto (\Lambda^2 \alpha,\Lambda^{-2}\beta)$ and fix one of them up to sign to be $\pm1$. The rescaling can be done without spoiling the gauge \eqref{eq:so11metricsemi}. The metric is always lorentzian and mostly plus\footnote{The signature can only change when $\det{B_2}=0$, equivalently $2\gamma=\alpha+\beta$, so we can check the signature for a few values of $\alpha,\beta,\gamma,\delta$.}.
 
If $\theta=0$, the automorphism group is enhanced to \begin{equation}\label{eq:auttheta0}
\mathrm{Aut}(\mathfrak{\mysosmall}(1,1;0))=
\mathrm{GL}(2,\RR)\ltimes \RR^{1,1}\end{equation}
and one can then fix the metric $B_1$ to be
\begin{equation}\label{eq:so11zeroB}
 B_1 = |\delta|\left(\pm \tilde{l}\,\tilde{l} \pm\,\tilde{m}_1\,\tilde{m}_1\pm \,\tilde{m}_2\,\tilde{m}_2\right)~,\quad\text{for }\theta=0 ~.
 \end{equation}
The metric is in fact maximally symmetric and the space is, according to the signs above: anti-de Sitter ($++-$), de Sitter ($-++$), or hyperbolic space ($+++$). Similarly, one can use the automorphisms in \eqref{eq:auttheta0} to bring the metric $B_2$ to the form
\begin{equation}
 B_2 = \delta \, \tilde{l}\,\tilde{l}+ \tilde{m}_1 \, \tilde{l} + \tilde{m}_2\,\tilde{m}_2~,\quad\text{for }\theta=0~.
\end{equation}
This metric turns out to be flat Minkowski.

Without further ado, we check the Einstein equation on \eqref{eq:so11metric} and \eqref{eq:so11metricsemi} in order to identify the maximally symmetric spaces. The metric $B_1$ is Einstein if and only if
\begin{itemize}
 \item $\theta=0$, in which case $R=-24/\delta$ (anti-de Sitter, hyperbolic or de Sitter).
 \item $\theta=\frac\pi4$ and $\alpha=0$, in which case $R=-12/\delta$ (anti-de Sitter),
 \item $\alpha=\beta=0$, in which case $R=-24\cos^2\theta/\delta$ (anti-de Sitter or flat).
\end{itemize}
The metric $B_2$ is Einstein if and only if 
\begin{itemize} \item $\theta=0$, or \item $\theta=\frac\pi4$ and $\alpha=0$.
\end{itemize}
In both of these last two cases, $B_2$ gives flat space.

\subsubsection*{Equations of Motion}
The next step to take is to check if and when the metric in \eqref{eq:so11metric} or \eqref{eq:so11metricsemi} can satisfy the Einstein-Cotton equation for a choice of parameters $(\theta,\mu,\delta,\alpha,\beta,\gamma)$. 
The space with metric $B_1$ is not Einstein but satisfies the Einstein-Cotton equation if and only if
\begin{ECsols}
 \item
 \label{item:so11warpedA}
 $\theta=\frac\pi4$ and $\alpha\beta\neq0$, in which case 
  \begin{align}
 R&=\frac{4}{\delta}\frac{4\alpha\beta-3\gamma^2}{\gamma^2-\alpha\beta}~,
 &
 \mu &= \frac{\sqrt{|\delta(\alpha\beta-\gamma^2)|}}{3\sqrt{2}\gamma}~.
 \end{align}
\item\label{item:so11ppA} $\alpha=0$, but the rest of the parameters do not reduce it to anti-de Sitter. The parameters that enter the equation are
  \begin{align}
 R&=-\frac{24}{\delta}\cos^2\theta~,
 &
 \mu &= \frac{\sqrt{\delta}}{2\left(\cos\theta-2\sin\theta\right)}~.
 \end{align} 
\item\label{item:so11ppB} $\beta=0$, but the rest of the parameters do not reduce it to anti-de Sitter. The parameters that enter the equation are
  \begin{align}
 R&=-\frac{24}{\delta}\cos^2\theta~,
 &
 \mu &= \frac{\sqrt{\delta}}{2\left(\cos\theta+2\sin\theta\right)}~.
 \end{align}
\end{ECsols}
Note that \ref{item:so11ppA} and \ref{item:so11ppB} are not isometric, see the footnote on page~\pageref{page:footZ}, provided as is generically the case that the isometry algebra does not enhance. These two metrics are also lorentzian, mostly plus for $\delta>0$ and so $R\leq0$. We now turn to the space with metric $B_2$. It satisfies the Einstein-Cotton equation and is not Einstein if and only if 
\begin{ECsols} 
\item $\theta=\frac\pi4$ and $\alpha\beta\neq0$ with
\begin{align}\label{eq:ec10Rmu}
 R &= \frac{64\beta\gamma^2}{(\beta-\gamma)^2}~, &
 \mu^2 R & = \frac{2}{9} \sign(\beta)
\end{align}
\label{item:so11warpedB}
\end{ECsols}
Note that we cannot have $\beta=\gamma$ in \eqref{eq:ec10Rmu}, because since $B_2$ is defined with $\gamma^2=\alpha\beta$, then $\gamma=\beta=\alpha$ gives a degenerate metric. Recall the two Einstein spaces with metric $B_2$ that we know: $\theta=0$ and $\theta=\frac\pi4$ with $\alpha=0$, which are furthermore flat. The metric $B_2$ with $\theta=\frac\pi4$ and $\beta=0$ solves instead the Cotton equation $C_{ab}=0$. 

We calculate the Segre-Petrov type of \ref{item:so11warpedA} and \ref{item:so11warpedB} and find that they are both of type D. By the type-D theorem, \ref{item:so11warpedA} is  spacelike or timelike warped anti-de Sitter because $R^{(2)}<0$ in \eqref{eq:typeDR2}. On the other hand \ref{item:so11warpedB} has $R^{(2)}=0$ and it is thus warped lorentzian flat. This should not come as a surprise, given the two embeddings
\begin{align}
\mathfrak{\mysosmall}(1,1;\frac\pi4)
=
\left(\RR\ltimes\RR\right)\oplus\RR
&\subset
{\mathfrak{sl}_2}\oplus\RR~,
\\
\left(\RR\ltimes\RR\right)\oplus\RR
&\subset
\mathfrak{so}(1,1)\ltimes\RR^{1,1}\oplus\RR
\end{align}
and the symmetries of warped anti-de Sitter and warped Minkowski.

On the other hand, the solutions \ref{item:so11ppA} and \ref{item:so11ppB} are both (generically) of spacetime type N. Neither of them can be null warped anti-de Sitter when $\theta\neq\frac\pi4$, nor any other metric on $\mathrm{SL}(2,\RR)$. Indeed, when $\theta\neq\frac\pi4$, $\mathfrak{\mysosmall}(1,1;\theta)$ is not a subalgebra of ${\mathfrak{sl}_2}\oplus \RR$. We identify these solutions with the pp-waves. 

\subsubsection*{Identifying \ref{item:so11ppA} and \ref{item:so11ppB}.}

The general TMG pp-wave metric in anti-de Sitter space was given in \eqref{eq:oldpp}. We will put coordinates on  \ref{item:so11ppA} and \ref{item:so11ppB} and relate them to \eqref{eq:oldpp}.

We choose the group representative 
\begin{equation}
 \mathcal{V}(x,y,w)=e^{x\,m_1+y\,m_2}e^{w\, l}\in\mathrm{\mysocapital}(1,1;\theta)
\end{equation} and the Maurer-Cartan one-forms  are
\begin{multline} \mathcal{V}^{-1}\grad\mathcal{V}
 = \grad w \,l + \grad{x} \left(  \cosh(2 w \sin\theta) m_1 -\sinh(2 w \sin\theta)m_2\right)e^{-2 w \cos\theta}\\
 +\grad y \left( \cosh(2 w \sin\theta)m_2 -\sinh(2 w \sin\theta)m_1\right)e^{-2 w \cos\theta}~.
\end{multline}
The metric \ref{item:so11ppA} is \begin{equation}B_1=B_0+\beta(m_1+m_2)^2~,\end{equation} where $B_0=\delta\, \tilde{l}\,\tilde{l} +2\gamma (-\tilde{m}_1\,\tilde{m}_1+\tilde{m}_2\,\tilde{m_2})$ gives anti-de Sitter or flat space, respcetively for $\theta\neq\frac\pi2$ and $\theta=\frac\pi2$. 
The spacetime metric in the chosen coordinates is
\begin{multline}\label{eq:ec6is} g=B_1(\mathcal{V}^{-1}\grad\mathcal{V},\mathcal{V}^{-1}\grad\mathcal{V})= \delta \, \grad w^2 + 2\gamma e^{-4 w \cos\theta}(\grad x^2-\grad y^2)\\ + 
 \beta \left( \grad x - \grad y \right)^2 e^{-4 w(\cos\theta-\sin\theta)}~.
\end{multline}
We need to take cases according to whether $\theta=\frac\pi2$ or $\theta\neq\frac\pi2$.

Let us assume first that $\theta\neq\frac\pi2$. If we define 
\begin{equation}\label{eq:z4pp}
z=\sqrt{\frac{\delta}{8\gamma\cos^2\theta}} e^{2w\cos\theta},\end{equation} then the $\beta=0$ metric reveals the familiar Poincar\'e coordinates of anti-de Sitter. 
The $\beta$-deformation \eqref{eq:ec6is} becomes
\begin{equation}\label{eq:newpp} g =
 \frac{\delta}{4\cos^2\theta}\frac{\grad z^2+\grad x^2-\grad y^2}{z^2} +\beta\, \left( \grad x - \grad y \right)^2 z^{-2\frac{\cos\theta-\sin\theta}{\cos\theta}} ~.
\end{equation}
The constant $\beta$ can be normalized up to sign. By using $z=e^{-\rho}$, the TMG pp-wave \eqref{eq:oldpp} (with the plus sign) and \eqref{eq:newpp} agree with $f(u)=\sign(\beta)$. Note that at $\theta=\frac\pi4$ the metric is just a rewriting of anti-de Sitter and the spacetime reduces from type N to type O.

Similarly, the metric \ref{item:so11ppB} yields the same anti-de Sitter metric in \eqref{eq:ec6is} but with the deformation
\begin{equation}
 \alpha \left( \grad x + \grad y \right)^2 e^{-4 w(\cos\theta+\sin\theta)} ~.
\end{equation}
When $\theta\neq\frac\pi2$, we change to the $z$ coordinate via \eqref{eq:z4pp} and \ref{item:so11ppB} becomes
\begin{equation}\label{eq:newerpp}
\begin{aligned}
 g=\frac{\delta}{4\cos^2\theta}\frac{\grad z^2+\grad x^2-\grad y^2}{z^2} + \alpha \, \left( \grad x + \grad y \right)^2 z^{-2\frac{\cos\theta+\sin\theta}{\cos\theta}} ~.
 \end{aligned}
 \end{equation}
Again, by using the values of $\mu$ and $R$ of this solution, we match with the pp-wave metric \eqref{eq:oldpp} (with plus sign) and $f_0=\pm1$. When $\theta=\frac\pi4$ this space is the null warped AdS with the enhanced isometry ${\mathfrak{sl}_2}\oplus\RR$. When $\theta=\frac\pi2$ the two metrics give
 \begin{equation}\label{eq:flatpp} g= \delta\, \grad w^2  +\grad x^2-\grad y^2 + 
 \beta' \left( \grad x \mp \grad y \right)^2 e^{\pm 4 w}~.
 \end{equation}
They are isometric by sending $w\mapsto-w$ and $y\mapsto-y$, which can be explained due to an enhanced $\ZZ_2\subset\mathrm{O}(1,1)$ automorphism of the Lie algebra. The metric \eqref{eq:flatpp} is the flat space pp-wave.

\subsubsection*{Extra isometries}
By using the Killing vector tests \eqref{eq:killingtest} and \eqref{eq:killingtest2}, we seek to find when the general metric has extra isometries. It turns out that $B_1$ can have extra isometries only if $\theta=\frac\pi4$. For simplicity we now restrict to the TMG solutions. The metric $B_1$ at $\theta=\frac\pi4$ gives one of the warped solutions with isometry ${\mathfrak{sl}_2}\oplus\RR$. For $\beta\neq0$ the space is warped spacelike or warped timelike and an extra Killing vector is $d=m_1-m_2$ with $B(d,d)=\beta$. For $\beta=0$ the space is null warped where \eqref{eq:newerpp} gives the metric
\begin{equation}
\frac{\delta}{2}\frac{\grad z^2+\grad x^2-\grad y^2}{z^2} + \frac{\left( \grad x + \grad y \right)^2}{ z^{4} } ~.
 \end{equation}
On the other hand, the $B_2$ solution is warped flat with enhanced symmetries.

\subsection{Metrics on \texorpdfstring{$\mathrm{\mysocapital}(2;\theta)$}{\mysocapital(2;theta)}}
\label{sec:so2theta}
We define the dual basis $\{\tilde{l},\tilde{m}_1,\tilde{m}_2\}$ to the Lie algebra basis $\{{l},{m}_1,{m}_2\}$ of $\mathfrak{\mysosmall}(2;\theta)$, where we recall the Lie brackets from \eqref{eq:commso2}:
\begin{align}
 [l,m_1]&= 2\cos\theta\, m_1 + 2\sin\theta\, m_2 ~,&
 [l,m_2]&= 2\cos\theta\, m_2 - 2\sin\theta\, m_1 .
\end{align}
A left-invariant metric on $\mathfrak{\mysosmall}(2;\theta)$ is given by a metric $B$ on $\mathfrak{\mysosmall}(2;\theta)$ up to the automorphism group. For $\theta\neq0$:
\begin{equation}\label{eq:autso2theta} \mathrm{Aut}(\mathfrak{\mysosmall}(2;\theta))=
(\mathrm{SO}(2)\times\RR^+)\ltimes \RR^2\text{ if } \theta\neq0~.
\end{equation}
The $\mathrm{SO}(2)$ automorphisms act on the $\{m_i\}\in\RR^2$ in the fundamental and can be used to diagonalize the metric, $B(m_1,m_2)=0$. The $\RR^2$ automorphisms send $l\mapsto l+x \, m_1 +y\, m_2 $. 
At $\theta=\frac\pi2$ we also have the automorphism that sends $(l,m_1,m_2)\mapsto(-l,m_2,m_1)$.


We have already covered $\theta=0$ in section \ref{sec:so11} and henceforth focus on $\theta\neq0$. With operations similar to those in section \ref{sec:so11}, but with $B(m_1,m_2)=0$, we fix the metric into two forms according to whether $B(m_i,m_j)$ is degenerate or not:
\begin{align} B_1 &= \alpha \,\tilde{l}\,\tilde{l} + \beta\, \tilde{m}_1\,\tilde{m}_1 +\gamma \,\tilde{m}_2\,\tilde{m}_2 ,\\\intertext{where we can furthermore rescale $\beta$ or $\gamma$ freely by using the $\RR^+$ automorphisms, and}
B_2 &= \alpha\, \tilde{l}\,\tilde{l}+\tilde{l}\, \tilde{m}_1 + \gamma \,\tilde{m}_2\,\tilde{m}_2 ~.
\end{align}
The metric $B_2$ is lorentzian, mostly plus for $\alpha>0$. 

Let us arrange for mostly plus lorentzian or euclidean signatures. The metric $B_1$ is Einstein if and only if
\begin{itemize}
 \item $\theta=0$. It is hyperbolic space, anti-de Sitter or de Sitter, depending on the signature and scalar curvature $R=-24/\alpha$.
 \item $\beta=\gamma$. It is hyperbolic space, de Sitter, or flat, with 
 \[R=-\frac{24\cos^2\theta}{\alpha}.\]
\end{itemize}
The metric $B_2$ is Einstein if and only if
\begin{itemize}
 \item $\theta=0$, in which case it is flat.
\end{itemize}
We then check if and when the metrics can satisfy the Einstein-Cotton equation for $\mu\neq0$. We find that there are no other solutions, whether we use $B_1$ or $B_2$. 

\subsection{Metrics on \texorpdfstring{$\mathrm{A}_\infty$}{a-infty}}
The Lie algebra $\mathfrak{a}_\infty$ is spanned by $r$, $x$ and $y$ and has non-vanishing bracket $[r,x]=-y$. We define the dual basis $\{\tilde{r},\tilde{x},\tilde{y}\}$. It is a limit of both $\mathfrak{\mysosmall}(2;\theta)$ and $\mathfrak{\mysosmall}(1,1;\theta)$, so it should come as no surprise that its automorphism group is larger than each:
\begin{equation}\mathrm{Aut}(\mathfrak{a}_\infty)
=
\left( \ZZ_2\times \mathrm{SL}(2,\RR)\times\RR^+\right)\ltimes \RR^2~. 
\end{equation}
The $\mathrm{SL}(2,\RR)$ automorphisms act on the vector $(r,x)$ in the fundamental, thus preserving the bracket 
$[r,x]=-y$, whereas $\ZZ_2\times\RR^+$ acts as $(r,x,y)\mapsto (r,\pm\Lambda\,x,\pm\Lambda y)$. Finally, the $\RR^2$ automorphisms send $(r,x,y)\mapsto (r+a\,y,x+b\,y,y)$. The $ \RR^2$ automorphisms act on the dual space as $\tilde{y}\mapsto \tilde{y}-a\,\tilde{r}-b\,\tilde{x}$ and by also using the rescalings in $\RR^+$, a metric can be fixed as
\begin{equation}
B = \pm\tilde{y}\,\tilde{y}+\alpha\,\tilde{r}\,\tilde{r}+2\gamma\,\tilde{r}\,\tilde{x}+\beta\,\tilde{x}\,\tilde{x}~.
\end{equation}
The $\mathrm{SO}(2)\subset\mathrm{SL}(2,\RR)$ automorphisms can be used to diagonalize the metric, $\gamma=0$, and we can furthermore rescale $\alpha$ or $\beta$ freely.

The metric $B$ is never Einstein, but satisfies the Einstein-Cotton equation 
\begin{ECsols}
\item 
\label{item:flatwarped}
always with
 \begin{align}
 \mu^2 &=\frac{4}{9}|\gamma^2-\alpha\beta|~, &
 R&=\pm\frac{1}{2(\gamma^2-\alpha\beta)}~.
 \end{align}
\end{ECsols}
The solution is type D and an application of \eqref{eq:typeDR2} shows it is warped flat. The  Baker-Campbell-Hausdorff  formula allows us to write a representative as $\mathcal{V}=e^{s\,r}e^{t\,x}e^{u\,y}$ and the metric space is manifestly a fibration over a flat space
\begin{equation}\label{eq:flatbase}
 g= B(\mathcal{V}^{-1}\grad \mathcal{V},\mathcal{V}^{-1}\grad \mathcal{V})
=
\pm\left(\grad u - t \,\grad s\right)^2+\alpha\, \grad s^2+2\gamma\,\grad s\,\grad t+\beta\,\grad t^2~.
\end{equation}
The full isometry algebra is thus an extension by $\partial_u$ of $\mathfrak{so}(2)\ltimes \RR^2 $ or $\mathfrak{so}(1,1)\ltimes\RR^{1,1}$ depending on the signature of the base space. Specifically, $\mathfrak{a}_\infty$ is spanned by the translations $\partial_s$, $\partial_t+s\,\partial_u$ and $\partial_u$. 
The metric can be rewritten, by absorbing most of the constants, as
\begin{equation}
g=\sigma_1\,\left(\grad u + \nu \, t \,\grad s\right)^2+\grad x^2+\sigma_2\,\grad t^2~,
\end{equation}
where the two signs $\sigma_i$ are uncorellated.

\subsection{Metrics on \texorpdfstring{$\mathrm{A}_0$}{a0}}
The Lie algebra $\mathfrak{a}_0$, spanned by $r$,$x$ and $y$, has non-vanishing brackets
\begin{align}
[r,x]&=x~,&[r,y]&=x+y~.
\end{align} 
We define the dual basis $\{\tilde{r},\tilde{x},\tilde{y}\}$. The automorphism group is four-dimensional and contains the $\ZZ_2$-rescalings $(x,y)\mapsto (\pm\Lambda\,x,\pm\Lambda\,y)$ and the inner automorphisms with non-trivial action
\begin{align}
e^{a\,x+b\,y} r e^{-a\,x-b\,y} &= r + (a+b)x+b\, y\\\intertext{and}
e^{t\,r}\begin{pmatrix}x\\y\end{pmatrix}e^{-t\,r} &= e^{t}\begin{pmatrix}1&0\\t&1\end{pmatrix}
\begin{pmatrix}x\\y\end{pmatrix}~.
\end{align}
The latter two act respectively on the dual vector space as \begin{align}
(\tilde{x},\tilde{y})&\mapsto (\tilde{x}-(a+b)\tilde{r},\tilde{y}-b\,\tilde{r})\label{eq:auta01} \\\intertext{and}
(\tilde{x},\tilde{y})&\mapsto (e^{-t}\tilde{x},e^{-t}\tilde{y}-t\,e^{-t}\,\tilde{x})\label{eq:auta02}~.\end{align}
We will use the automorphisms to fix the metric into four classes.

Assume a general metric on $\mathfrak{a}_0$
\begin{equation}
B = \delta\, \tilde{r}^2 + 2\, b_1\, \tilde{r}\,\tilde{x} + 2
\,b_2\, \tilde{r}\,\tilde{y}
+ \alpha \,\tilde{x}^2 + 2\,\gamma \,\tilde{x}\,\tilde{y} + \beta\,\tilde{y}^2 ~.
\end{equation}
Say $\beta\neq 0$. Then we can use \eqref{eq:auta02} to set $\gamma=0$. If additionally $\alpha\neq0$ we can use \eqref{eq:auta01} to set $b_1=b_2=0$, otherwise we can use \eqref{eq:auta01} and the rescalings to set $b_1=1$ and $b_2=0$. Say $\beta=0$. If additionally $\gamma\neq0$ then we can use \eqref{eq:auta02} to set $\alpha=0$ and \eqref{eq:auta01} to set $b_1=b_2=0$, otherwise we can use the automorphisms to set $b_1=0$ and $b_2=1$. All together, the metric can be brought to one of the following forms
\begin{align}
B_1 &= \delta \,\tilde{r}^2 \pm \tilde{x}^2 + \beta \, \tilde{y}^2 \\
B_2 & = \delta\, \tilde{r}^2 \pm 2\,\tilde{x}\,\tilde{y} \\
B_3 & = \delta \, \tilde{r}^2 +\tilde{r}\,\tilde{x}+\beta \, \tilde{y}^2 \\
B_4 & = \delta \, \tilde{r}^2 + \tilde{r}\,\tilde{y} + \alpha \,\tilde{x}^2~.
\end{align}
In these metrics, we have used all of the automorphisms to fix the metric. 

The only metric that can be Einstein for a choice of parameters is $B_3$. In particular, $B_3$ is flat for all values of $\beta\delta\neq0$. Apart from this solution, the only metric that can satisfy the Einstein-Cotton equation is
\begin{ECsols}
\item \label{item:a0pp} the metric $B_2$ for all values of $\delta$, with $\mu^2=|\delta|$ and $R=-6/\delta$.
\end{ECsols}
The space is type $N$, but its isometry algebra does not enhance. In particular, the null vector $d_a$ fails the second test \eqref{eq:killingtest2}. 

Let us choose the representative $\mathcal{V}=e^{w\,x+u\,y}e^{s\,r}$ that can always be achieved by use of the Baker-Campbell-Hausdorff formula. We find the left-invariant Maurer-Cartan one-forms
\begin{equation}
\mathcal{V}^{-1}\grad\mathcal{V}= \left(\grad w \,e^{-s}-s\,e^{-s}\grad u\right)\,x +\grad u \,e^{-s}\,y+\grad s\,r~.
\end{equation}
The solution \ref{item:a0pp} in these coordinates is
\begin{equation}\label{eq:muellppg}
g= B_2(\mathcal{V}^{-1}\grad\mathcal{V},\mathcal{V}^{-1}\grad\mathcal{V})=\delta\,\grad s^2 \pm 2\,e^{-2\,s}\left(\grad w-s\grad u\right)\grad u~.
\end{equation}
We identify \eqref{eq:muellppg} with one of the two  $|\mu|=\ell$ pp-waves, see \eqref{eq:ppcrit1}. With this, we have finished the classification of homogeneous anisotropic TMG solutions that we summed in table~\ref{table:solutions}.

\section{Lie algebra deformations}\label{sec:LieCoho}
In this section we introduce the notion of infinitesimal Lie algebra deformations. We then describe the continuous deformations of three-dimensional Lie algebras. In the first part, we discuss the Lie algebra cohomology that describes Lie algebra deformations in some detail. However, the cohomological calculations that are needed in section \ref{sec:HomoDef} are deferred to the appendix \ref{app:coho}.

\subsection{Cohomology}
Consider a Lie algebra $\mathfrak{g}$ with Lie bracket $[-,-]$. A finite deformation $\mathfrak{g}_t$ of $\mathfrak{g}$ is an antisymmetric form $[-,-]_t:\Lambda^2\mathfrak{g}\rightarrow \mathfrak{g}$, analytic at $t=0$ where it reduces to the Lie bracket $[-,-]$, 
\begin{equation}\label{eq:finiteexpansion}
[-,-]_t = [-,-] + t\,f(-,-) +\sum_{n=2}^\infty \frac{t^n}{n!} f_{[n]}(-,-)~,
\end{equation}
and such that it satisfies the Jacobi identity for the whole domain of $t$. The deformation is trivial if the Lie algebra $\mathfrak{g}_t$ with bracket $[-,-]_t$ is isomorphic to the original. We call the Lie algebra rigid if it does not admit non-trivial finite deformations. 

The term $f:\Lambda^2\mathfrak{g}\rightarrow \mathfrak{g}$ in \eqref{eq:finiteexpansion} is called an infinitesimal deformation and the Jacobi identity to first order in $t$ is
\begin{equation}\label{eq:infJac} 0=f(a,[b,c])+[a,f(b,c)]+\text{cycl.} =: \grad f(a,b,c) .\end{equation}
If there is a linear transformation $g\in \mathfrak{gl}(\mathfrak{g})$ such that
\begin{equation}\label{eq:infTrivial} f(a,b) = -g([a,b])+[g(a),b]+[a,g(b)]  =: \grad g(a,b)~,\end{equation}
then the infinitesimal deformation is infinitesimally trivial. Not all non-trivial infinitesimal deformations can be integrated to finite deformations. The obstructions are given order by order in the Jacobi identity. We now reinterpet \eqref{eq:infJac} and \eqref{eq:infTrivial} in terms of cohomology, where the meaning of $\grad f$ and $\grad g$ will become clear. 
 
Assume a $\mathfrak{g}$-module $\mathfrak{m}$ and the space $C^{p}(\mathfrak{g},\mathfrak{m})$ of linear maps $\Lambda^p\mathfrak{g}\rightarrow\mathfrak{m}$. Let 
$m\in \mathfrak{m}$, $\alpha\in\mathfrak{g}^\ast$, $f\in C^{p}(\mathfrak{g},\mathfrak{m})$ and $a,b\in\mathfrak{g}$. Define a differential $\grad:C^{p}(\mathfrak{g},\mathfrak{m})\rightarrow C^{p+1}(\mathfrak{g},\mathfrak{m})$ by
\begin{align}
\grad\alpha(a,b) &= -\alpha([a,b]) \\
\grad m (a) &=  a\cdot m
\\\intertext{and extend it linearly} \grad( \alpha\wedge f)&=\grad\alpha\wedge f - \alpha\wedge \grad f~.
\end{align}
The cohomology of the complexes $C^{p}(\mathfrak{g},\mathfrak{m})$ with respect to $\grad$ is
\[ H^{p}(\mathfrak{g},\mathfrak{m})=\frac{Z^p(\mathfrak{g},\mathfrak{m})}{B^p(\mathfrak{g},\mathfrak{m})} =\frac{ \textrm{ker}\, \grad:C^{p}(\mathfrak{g},\mathfrak{m})\rightarrow C^{p+1}(\mathfrak{g},\mathfrak{m}) }{\textrm{im}\, \grad:C^{p-1}(\mathfrak{g},\mathfrak{m})\rightarrow C^{p}(\mathfrak{g},\mathfrak{m}) }~.\]
One can take the module $\mathfrak{m}=\mathfrak{g}$ in the adjoint representation. The infinitesimal non-trivial deformations \eqref{eq:infJac} are cocycles, $\grad f=0$, in $C^2(\mathfrak{g},\mathfrak{g})$ modulo the coboundaries \eqref{eq:infTrivial}, $f=\grad g$ with $g\in C^1(\mathfrak{g},\mathfrak{g})$. That is, infinitesimal non-trivial deformations make up the second cohomology $H^2(\mathfrak{g},\mathfrak{g})$. 

A finite deformation yields an infintesimal deformation in $H^2(\mathfrak{g},\mathfrak{g})$. Conversely, the obstruction to integrating an infinitesimal deformation is given order by order by elements in $H^3(\mathfrak{g},\mathfrak{g})$. Therefore, if $H^2(\mathfrak{g},\mathfrak{g})=0$ then the Lie algebra is rigid, whereas if $H^2(\mathfrak{g},\mathfrak{g})\neq0$ and $H^3(\mathfrak{g},\mathfrak{g})=0$ then the Lie algebra is not rigid. The same still holds if the term $f$ in \eqref{eq:finiteexpansion} is missing and we start with finite deformations of order two or higher. Finally, once we have the infinitesimal deformations in $H^2(\mathfrak{g},\mathfrak{g})$, we need to consider the action of the automorphisms $\mathfrak{aut}(\mathfrak{g})$ on $H^2(\mathfrak{g},\mathfrak{g})$, which still give isomorphic Lie algebras. A semisimple Lie algebra has  $H^2(\mathfrak{g},\mathfrak{g})=0$ and so semisimple Lie algebras are rigid. 

We will slightly refine the definition of deformations. If a Lie algebra $\mathfrak{g}_0$ can deform finitely to $\mathfrak{g}_t$ we will write it as $\mathfrak{g}_0\leadsto\mathfrak{g}_t$. If a deformation $[-,-]_t$ gives non-isomorphic algebras for all $t\in I$, then we say that the Lie algebra $\mathfrak{g}_0$ deforms into the family $\mathfrak{g}_t$, where we can be explicit and specify the domain $I$ of the parameter $t$. We write this as $\mathfrak{g}_0 \,\famdeforms \,\mathfrak{g}_t$, $t\in I$. One can literally interpret this as the Lie algebra $\mathfrak{g}_0$ being on the boundary of $\mathfrak{g}_t$, $t\in I$. More precisely, think of the algebraic curve given by the Jacobi identity on the structure coefficients. The curve is fibered over the Lie algebras by the action of linear transformations, and the structure coefficients for $\mathfrak{g}_0$ are on the boundary of those for $\mathfrak{g}_t$. 

Consider the case when the deformation $[-,-]_t$ gives isomorphic Lie algebras 
for all $t>0$ but for $t>0$ it is not isomorphic with the undeformed $[-,-]$. We say that the Lie algebra with bracket $[-,-]_1$ contracts to the undeformed algebra with bracket $[-,-]$ or, conversely, that the Lie algebra with bracket $[-,-]$ implodes into $[-,-]_1$. We will write this as $\mathfrak{g}_0\, \circright \,\mathfrak{g}_1$. We have hitherto introduced the notations
\begin{align}
\mathfrak{g}_0&\leadsto\mathfrak{g}_t \\
\mathfrak{g}_0 & \,\famdeforms \,\mathfrak{g}_t, \,t\in I\\
\mathfrak{g}_0 &\, \circright \,\mathfrak{g}_1
\end{align}
These are useful to expose the nature of the various deformations that we come across.

\subsection{Three-dimensional Lie algebra deformations}
The deformations of three-dimensional Lie algebras were studied in \cite{levy-nahas:deform} by using their cohomology. We will present here the deformations without proof that these exhaust all of them. In writing these, it has been instrumental to compare with the simple contractions in~\cite{conatser:196} and the generalized contractions in~\cite{weimar-woods:2028}, see also \cite{fialowski}. 

Firstly, the Lie algebras $\mathfrak{\mysosmall}(2;0)$ and $\mathfrak{a}_\lambda$ can only deform into the family $\mathfrak{a}_\lambda$. The deformation of $\mathfrak{a}_\lambda$ is obviously achieved by considering the family $\mathfrak{a}_\lambda$ with $\lambda\in\RR$. In particular, by considering the isomorphisms
\begin{align}
\mathfrak{\mysosmall}(2;\theta) &= \mathfrak{a}_\lambda~
\text{ with }\lambda=\tan^2\theta,
&
\theta\in(0,\frac\pi2)&\Leftrightarrow\lambda>0~,\\
\mathfrak{\mysosmall}(1,1;\theta)
&= \mathfrak{a}_\lambda~
\text{ with }
\lambda=-\tan^2\theta,
&\theta\in(0,\frac\pi2)&\Leftrightarrow\lambda<0~.
\end{align}
the $\mathfrak{\mysosmall}(2;\theta)$ and $\mathfrak{\mysosmall}(1,1;\theta)$ with $\theta\neq0,\frac\pi2$ can be deformed into the family $\mathfrak{a}_\lambda$ by varying $\lambda$. The Lie algebras $\mathfrak{\mysosmall}(2;\frac\pi2)$ and $\mathfrak{\mysosmall}(1,1;\frac\pi2)$ can similarly deform into the families $\mathfrak{\mysosmall}(2;\theta)$ and $\mathfrak{\mysosmall}(1,1;\theta)$ by varying $\theta$. The algebra $\mathfrak{\mysosmall}(2;0)=\mathfrak{\mysosmall}(1,1;0)$ can also deform into our families by varying $\theta$. The Lie algebra $\mathfrak{\mysosmall}(2;0)$ can also directly implode into $\mathfrak{a}_0$, $\mathfrak{\mysosmall}(2;0)\,\implodes\,\mathfrak{a}_0$, for instance by the deformation
\begin{align}
 [r,x]&= x \,,& [r,y]&= t \,x+y~.
\end{align}
Indeed, for $t\neq0$ the Lie algebra is isomorphic to $\mathfrak{a}_0$, e.g. rescale $x$ by $t$.

The Lie algebra $\mathfrak{a}_\infty$ can deform into the family $\mathfrak{a}_\lambda$ for arbitrarily large $\lambda$ in the manner that it was defined. However, $\mathfrak{a}_\infty$ can implode into the family $\mathfrak{a}_{\lambda}$ at any value of $\lambda$, e.g. the algebra
\begin{align}
 [r,x]&=t\,x-t^2\,\lambda\, y~, & [r,y]&=x+t\,y~,
\end{align}
is $\mathfrak{a}_{\lambda}$ by rescaling $r'=1/t\,r$ and $y'=t\,y$. In this sense, it can also deform into $\mathfrak{a}_{\lambda}$ for arbitrary small $\lambda$. We will represent these deformations as $\mathfrak{a}_\infty\implodes \mathfrak{a}_0$, $\mathfrak{a}_\infty\implodes \mathfrak{\mysosmall}(2;\theta)$ and $\mathfrak{a}_\infty\implodes \mathfrak{\mysosmall}(1,1;\theta)$ with $\theta\in(0,\frac\pi2)$.

The algebra $\mathfrak{a}_\infty$ can also implode into $\mathfrak{\mysosmall}(1,1;\frac\pi2)$ and $\mathfrak{\mysosmall}(2;\frac\pi2)$. For instance, $\mathfrak{a}_\infty\,\implodes\,\mathfrak{\mysosmall}(2;\frac\pi2) $ and $\mathfrak{a}_\infty\,\implodes\,\mathfrak{\mysosmall}(1,1;\frac\pi2) $ are given by
\begin{align}
 [r,x]&= -y\,, & [r,y]&=  t\,x\, ,
\end{align}
for respectively $t<0$ and $t>0$. The Lie algebras  $\mathfrak{a}_\infty$, $\mathfrak{\mysosmall}(1,1;\frac\pi2)$ and $\mathfrak{\mysosmall}(2;\frac\pi2)$ can also implode into the simple Lie algebra $\mathfrak{sl}_2$. For instance, $\mathfrak{\mysosmall}(1,1;\frac\pi2)\implodes\mathfrak{sl}_2$ is given by
\begin{equation}\begin{aligned}[]{}
[\tau_1,\tau_0] &= - t\, \tau_2~,\\
[\tau_1,\tau_2] &= - \tau_0~,\\
[\tau_0,\tau_2] &= - \tau_1~,
\end{aligned}\label{eq:sl2TOsoR}
\end{equation}
with $t>0$. The contraction $\mathfrak{a}_\infty\implodes\mathfrak{sl}_2$ can be defined by replacing the second of the above brackets with $[\tau_1,\tau_2] = - t\,\tau_0$. The contraction $\mathfrak{\mysosmall}(2;\frac\pi2)\implodes\mathfrak{sl}_2$ can be written as
\begin{equation}\begin{aligned}[]{}
[\tau_1,\tau_0] &= -  \tau_2~,\\
[\tau_1,\tau_2] &= - t\, \tau_0~,\\
[\tau_0,\tau_2] &= - \tau_1~,
\end{aligned}
\end{equation}
with $t>0$. Similarly, the Lie algebras $\mathfrak{a}_\infty$ and $\mathfrak{\mysosmall}(2;\frac\pi2)$ can also implode into the simple Lie algebra $\mathfrak{su}_2$, by inserting a power of $t$ on the right-hand side of (two or one respectively) $\mathfrak{su}_2$ brackets. However, $\mathfrak{\mysosmall}(1,1;\frac\pi2)$ cannot implode into $\mathfrak{su}_2$, see also \cite{weimar-woods:2028}.

In turn, the simple Lie algebras cannot be infinitesimally deformed because simple algebras are rigid. Finally, all Lie algebras can contract to the abelian Lie algebra. Indeed, take any basis $\tau_a$ of a Lie algebra $\mathfrak{g}$, and use the rescaled basis $t\,\tau_a$. At $t\neq0$ this is just a rescaling, but at $t\rightarrow0$ the Lie algebra contracts to the abelian one.

\begin{figure}[tbh]
\begin{center}
\tikzstyle{vertex}=[draw]
\tikzstyle{edgefam} = [o-, shorten <=3pt , shorten >=3pt ]
\tikzstyle{edgeimplodes} = [o-latex, shorten <=3pt, shorten >=3pt]
\begin{tikzpicture}[scale=1.3,auto,swap]
\foreach \pos/\name/\group in {
	{(1,6)/a/$\mathfrak{su}_2$}, 
	{ (-1,6)/i/$\mathfrak{sl}_2$ },
	{(0,4)/b/$\mathfrak{a}_\infty$}, 
	{(-3,4)/c/$\mathfrak{\mysosmall}(1,1;\frac\pi2)$},
	{(-3,2)/d/$\begin{array}{r}\mathfrak{\mysosmall}(1,1;\theta),\\{\theta\in(0,\frac\pi2)}\end{array}$}, 
	{(0,0)/e/$\mathfrak{\mysosmall}(2;0)$},
	{(0,2)/f/$\mathfrak{a}_0$}, 
        {(3,2)/g/$\begin{array}{r}\mathfrak{\mysosmall}(2;\theta),\\{\theta\in(0,\frac\pi2)}\end{array}$},
	{(3,4)/h/$\mathfrak{\mysosmall}(2;\frac\pi2)$}}{
	\node[vertex] (\name) at \pos {\group};}
\foreach \source/\dest  in {b/a, b/c, e/f, h/a, b/c, b/h, b/i, c/i, h/i, b/f, b/d,b/g}{
\draw[edgeimplodes] (\source) -- (\dest);}
\foreach \source/ \dest in {c/d, e/d, f/d, e/g, f/g, h/g, b/a}{
\draw[edgefam] (\source) -- (\dest);}
 \path[every loop/.append style={o-}] 
 	 (d) edge [loop left] () 
 	 (g) edge [loop right] ();
\end{tikzpicture}
\caption{Deformations of three-dimensional Lie algebras.}\label{fig:def3d}
\end{center}
\end{figure}
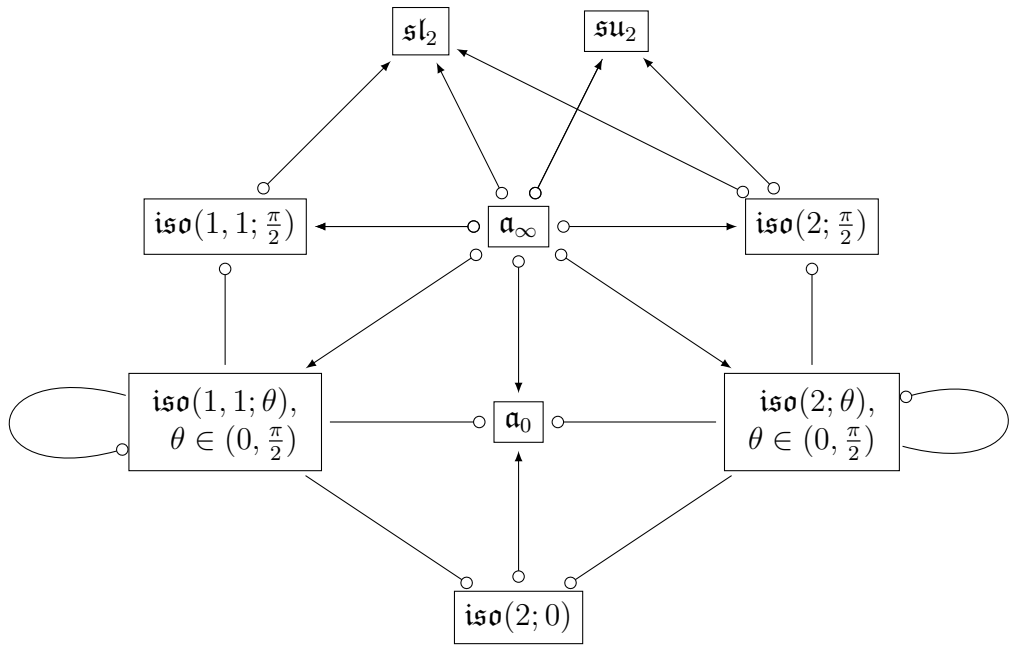

We show all deformations, except for the abelian contraction, in figure \ref{fig:def3d}. In general, one can achieve a finite deformation from a non-simple three-dimensional Lie algebra to any other non-simple Lie algebra by deforming the action $\rho_{ij}$ of $l$ on its nilpotent two-dimensional subagebra, $[l,m_i]=\sum_{j=1}^2\rho_{ij}m_j$. A finite deformation yields an infinitesimal deformation and it is this information of locality that is described in figure \ref{fig:def3d}. The contractions in the figure, that is the inverse of imploding $A\implodes B$, are written as (generalized) In\"on\"u-Wigner contractions in \cite{weimar-woods:2028}.

\section{Homogeneous deformations}\label{sec:HomoDef}
A homogeneous spacetime $(G/H,B)$ can be deformed by smoothly varying the group structure $G$ and $H$ or the metric $B$ at a fixed point, or both. In this section we consider homogeneous deformations of the solutions we have encountered, such that the deformation still solves TMG. We allow $R$ and $\mu$ in the equations of motion to vary with the deformation. However, we are interested in the deformations that are not related by a homothety. By a homothety transformation and change of orientation one can always fix $\mu=1$ and allow only $R$ to vary with the deformation. We are also not interested in the abelian contraction that gives flat space. If there is a three-dimensional transitive group $G$, we can choose $H=1$ and restrict to the anisotropic deformations within the solutions of section \ref{sec:3dGeometry}. We first describe in \S\ref{sec:anisodef} the anisotropic homogeneous deformations of TMG within the class of these solutions.

The isometry group of the anisotropic solutions has dimension larger than three for the type D spacetimes, the null warped anti-de Sitter, and the maximally symmetric spaces. These spacetimes can thus be described as isotropic cosets $(G/H,B)$. In order to find the homogeneous deformations of all anisotropic solutions, for each $\hat{G}/\hat{H}$ spacetime of our solutions, with $\hat{H}\neq1$ and $\hat{G}$ the full isometry group, we find the transitive subroups $G$ in $\hat{G}$ that have dimension larger than three. A deformation can be in varying the group structure of $G$ and its isotropy $H\subset{G}$, varying the metric at a fixed point in $G/H$, or both. All pairs $(G,H)$ have to be considered. After having described the $H=1$ deformations in \S\ref{sec:anisodef}, we show in the rest of the text that all $(G,H,B)$ \emph{infinitesimal} homogeneous deformations  are within the class of anisotropic deformations. Furthermore, we show that among maximally symmetric spaces, only anti-de Sitter, the sphere 
and flat space can be homogeneously deformed in TMG. We conclude with how one can obtain from the anisotropic solutions with a limiting procedure a non-trivial isotropic space $G/H$ that is not a group manifold.

\subsection{Anisotropic deformations}\label{sec:anisodef}
The simplest type of a homogeneous deformation of an anisotropic solution $(G,B)$ is where we keep the group $G$ fixed and vary smoothly the metric $B$ at the identity. It is worth describing these deformations first, before allowing the three-dimensional group to deform. In the calculations of section \ref{sec:3dGeometry}, we have fixed the metric $B$ under the action of the automorphisms of the group $G$. One can smoothly vary $B$ so that it crosses over these types, but this is not so clear once we have fixed the metric. 

Take for example that of warped anti-de Sitter. We can write it as a deformation from anti-de Sitter,
\begin{equation}\label{eq:warpedd}
\begin{aligned}
 g_w &= L\left( -\theta^0\,\theta^0+\theta^1\,\theta^1+\theta^2\,\theta^2\right) +t\, \left(d_a\,\theta^a\right)^2 ~.
\end{aligned}
\end{equation}
The vector $d_a$ can be timelike, spacelike or null with respect to the anti-de Sitter metric $\eta_{ab}$. Previously we had fixed $d_a\theta^a$ with a $\mathrm{SO}(1,2)$ rotation to be proportional to one of $\theta^0$, $\theta^2$, $\theta^0\pm\theta^1$. We have also included a scalar deformation $t$ to capture a sign that cannot be absorbed in $d_a$. By varying the norm of $d_a$ and $t$, one can cross over the metrics of timelike, spacelike or null anti-de Sitter. In the form where $B$ is fixed as in section \ref{sec:3dGeometry}, this is not so clear: crossing over from, say, spacelike warped to null warped anti-de Sitter would require an infinite boost of a spacelike $d_a$ while keeping its norm finite.

In \eqref{eq:warpedd} we have included the possibility that $L<0$. We then require that the space is mostly plus lorentzian, which is still possible with a spacelike vector $d^a$ provided that $t\,\eta_{ab}d^a d^b>|L|$. This corresponds to taking a negative $\beta=-\alpha$ in the spacelike warped metric. The two metrics $L>0$ and $L<0$ in \eqref{eq:warpedd} appear disconnected, but they are connected by $L=0$ and $t=0$ through a contraction of the Lie algebra $\mathfrak{a}_\infty\implodes{\mathfrak{sl}_2}$. This gives the warped flat space from a limit of spaclike or timelike anti-de Sitter. In all $L\neq0$ cases, the metric \eqref{eq:warpedd} is non-degenerate for $t\,\eta_{ab} d^a d^b \neq -L$. 

Let us fix the Einstein-Cotton parameter to be $\mu=1$. This can always be achieved with a metric homothety and a change of orientation. It is then useful to consider the dimensionality of these families of solutions. When doing so, $R$ can be a free parameter or not, depending on the family of solutions. For instance, we have already unified the spacelike, timelike and null warped anti-de Sitter in a one-dimensional family in \eqref{eq:warpedd}. Similarly, one can combine all the $R=0$ solutions on $\mathrm{SL}(2,\RR)$ into one family that is generically of type I. 

Indeed, if we solve the $R=0$ Einstein-Cotton equation on the $\mathrm{SL}(2,\RR)$-invariant metric
\begin{subequations}\label{eq:R0unified}
\begin{equation}\label{eq:R0unified1}
 g=\alpha \, \theta^0\,\theta^0+2\delta\,\theta^0\,\theta^1+ \beta\,\theta^1\,\theta^1+\gamma\,\theta^2\,\theta^2 ~.
\end{equation}
The solution is for 
\begin{equation}\label{eq:R0unified2}
(\alpha+\beta+\gamma)^2=4\beta\gamma+4\delta^2~.\end{equation}\end{subequations}
 It is locally one-dimensional if we also consider
\begin{itemize}
 \item the action of the automorphisms in $\mathrm{SO}(1,1)$ acting on the vector $(\theta^0,\theta^1)$,
 \item the condition $\mu=1$.
\end{itemize}
The solution reduces to any of the three $R=0$ solutions \ref{item:sl2triaxial}, \ref{item:sl2CSIKundt} and \ref{item:sl2typeIC} by an automorphism. Which one it reduces to depends on the sign of the determinant $(\alpha+\beta)^2-4\delta^2$. 

We have thus described the warped anti-de Sitter and the $R=0$ solutions as two separate one-dimensional families. The two families furthermore intersect. Spacelike or timelike warped anti-de Sitter at $R=0$ matches with the ``triaxial'' $R=0$ solution \ref{item:sl2triaxial} at two separate solutions that were found by Vuorio~\cite{Vuorio:1985ta}. Something equivalent to this happens for the $\mathrm{SU}(2)$-invariant metrics. The one-dimensional family of squashed / stretched sphere \ref{item:su2warped}, at $R=0$ and euclidean signature, matches the triaxial $R=0$ deformed sphere \ref{item:su2triaxial} at the euclidean Vuorio solution~\cite{Vuorio:1985ta}.

Next in complexity, we discuss varying both the group structure of the three-dimensional Lie groups and the metric on them. The infinitesimal deformations of the groups are summarized simply in figure \ref{fig:def3d}. 
It is important to note that, when deforming the Lie algebra strucure we keep the underlying vector space fixed. The metric $B$ on the vector space should thus vary smoothly. A more powerful restriction than comparing the form of the metric $B$ as we vary $G$ is comparing the scalar invariants of the solutions. We therefore fill figure~\ref{fig:def3d} with solutions and compare the smoothness of the invariants $R$, $\mu$, or $\mu^2 R$. We will first consider the TMG pp-waves on $\mathrm{\mysocapital}(1,1;\theta)$ or $\mathrm{A}_0$, the warped flat space \ref{item:flatwarped} on $\mathrm{A}_\infty$, and the flat space pp-wave on $\mathrm{\mysocapital}(1,1;\frac\pi2)$. 
With the exception of these solutions and the deformed sphere, all other solutions we came across were Einstein: anti-de Sitter, the sphere, de Sitter, hyperbolic space, and flat space. 

In general, anti-de Sitter can deform into the warped anti-de Sitter or its pp-waves. Null warped anti-de Sitter can deform to other pp-waves by varying $\mathfrak{\mysosmall}(1,1;\theta)$ from $\theta=\frac\pi4$. The pp-waves can limit to the $|\mu|=\ell$, where the isometry is deformed as $\mathfrak{a}_0\famdeforms\mathfrak{\mysosmall}(1,1;\theta)$. The coordinate description of this limit was described in \cite{Gibbons:2008vi}. The flat spacetime likewise deforms to warped flat space or its pp-wave. The warped flat spacetime can also implode into spacelike or timelike warped anti-de Sitter, whereby $\mathfrak{a}_\infty\implodes{\mathfrak{sl}_2}$. This latter contraction involves the flat $R^{(2)}\rightarrow 0$ limit of the two-dimensional base space, which is always easy to arrange for in a coordinate manner. 

Another non-trivial contraction is  ${\mathfrak{\mysosmall}(1,1;\frac\pi2)}\implodes{\mathfrak{sl}_2}$, in which the $R=0$ family on $\mathfrak{sl}_2$ can contract to the flat pp-wave. This is easy to describe by following the  Lie algebra contraction in \eqref{eq:sl2TOsoR}. We first rescale $\theta^i\mapsto \sqrt{t} \, \theta^i$ for $i=0,1$ in the general $R=0$ solution of \eqref{eq:R0unified1}. We then rescale the parameters $\alpha,\beta,\delta$ in \eqref{eq:R0unified1} by a factor of $1/t$ so that the $B_{ab}$ are finite as $t\rightarrow0$. Finally, \eqref{eq:R0unified2} gives the solution $(\alpha+\beta)^2=4\delta^2$ in the limit. This is precisely the form of the pp-waves \ref{item:so11ppA},
 \begin{equation}
  B_{1} = \gamma \,\tilde{l}\,\tilde{l} + 
 \frac{\alpha+\beta}{2}\,\left(\tilde{m}_1\pm \tilde{m}_2\right)^2+ \frac{\alpha-\beta}{2}\,\left(\tilde{m}_1\,\tilde{m}_1-\tilde{m}_2\,\tilde{m}_2\right) ~.
 \end{equation}
At the same time, the isometry algebra has contracted to $\mathfrak{\mysosmall}(1,1;\frac\pi2)$. We also have the similar deformations of the deformed sphere solutions of either signature, either into warped flat space or the flat pp-wave. Other deformations, like the warped flat space to pp-waves or the type D warped anti-de Sitter to the flat pp-wave, are not possible as we verify by comparing their scalar invariants. 

The two maximally symmetric spaces of de Sitter and hyperbolic space cannot be infinitesimally deformed in the space of anisotropic solutions. Indeed, first note that they only appear as solutions on $\mathrm{\mysocapital}(2;\theta)$ with $\theta\in[0,\frac\pi2)$. They cannot deform as solutions on $\mathrm{\mysocapital}(2;0)$ into the pp-waves on $\mathrm{\mysocapital}(1,1;\theta)$ or implode into the pp-wave on $\mathrm{A}_0$, see figure~\ref{fig:def3d}, because the signs of the scalar curvature and their signature do not match with the latter. As a family on $\mathrm{\mysocapital}(2;\theta)$ they can only limit to flat space as a solution on $\mathrm{\mysocapital}(2;\frac\pi2)$. It remains to show that they cannot limit to the warped flat space on $A_\infty$, $\mathfrak{a}_\infty \implodes \mathfrak{\mysosmall}(2;\theta)$. An argument against such a contraction is Geroch's result that the dimension of the isometry algebra cannot reduce in a spacetime limit \cite{Geroch:1969ca}. 
However, they can possibly homogeneously deform by considering them as homogeneous spaces with isotropy $G/H$, $H\neq1$. This is also potentially true for the other anisotropic solutions wherever the full isometry algebra is larger than dimension three. We discuss these possibilities in the following subsections.

\subsection{Anti-de Sitter}\label{sec:so22}
The isometries of anti-de Sitter are an appropriate cover of the symmetries of its quadric on $\RR^{2,2}$:
\[\mathrm{SO}(2,2)=\left( \mathrm{SL}(2,\RR)_L\times\mathrm{SL}(2,\RR)_R \right)/{\ZZ_2}~.\] Consider a homogeneous deformation of anti-de Sitter, $G_t/H_t$ with a $\mathfrak{h}_t$-invariant metric $B_t$ on $\mathfrak{g}_t/\mathfrak{h}_t$ for each $t$. At $t=0$ we must have 
\[ \mathfrak{g}_0\subseteq {\mathfrak{sl}_2}_L\oplus{\mathfrak{sl}_2}_R~.\]
Furthermore, $\mathfrak{g}_0$ should act transitively on anti-de Sitter. 

We find the suitable subalgebras in the appendix \ref{app:goursatcosets} by using Goursat's lemma. As it turns out, the transitive subalgebras are
\begin{align}
\mathfrak{g}_0 &= {\mathfrak{sl}_2}\oplus \mathfrak{h}
\\\intertext{with $\mathfrak{h}$ one of $0$, $\RR$, $\RR\ltimes\RR$, or ${\mathfrak{sl}_2}$, or}
\mathfrak{g}_0 &= \mathfrak{\mysosmall}(1,1;\theta)~\text{ with }\theta\neq\frac\pi2~.\end{align}
We have already defined the bases of the left and right action generators $l_a$ and $r_a$, which are orthonormal with respect to the Killing form. The Lie algebra of $\mathfrak{\mysosmall}(1,1;\theta)$ is generated up to conjugacy by $r_2+\mu \,l_2$, $l_0+l_1$ and $r_0+r_1$, where $\tan\theta=\left|\frac{1-\mu}{1+\mu}\right|$. We simply note here that $\mu\neq-1$ because otherwise the subalgebra is not transitive. By the factorization theorem in appendix~\ref{app:coho}, the algebras ${\mathfrak{sl}_2}\oplus \mathfrak{h}$ are rigid under deformations. In this case we have $\mathfrak{g}_t=\mathfrak{g}_0$ for all values of $t$. On the other hand, $\mathfrak{\mysosmall}(1,1;\theta)$ can be deformed as described in section \ref{sec:anisodef}. 

Let us assume that for some value of $t$, the subalgebra ${\mathfrak{sl}_2}$ in $\mathfrak{g}_t={\mathfrak{sl}_2}\oplus\mathfrak{h}$ does not act transitively on the deformed space and so, the deformation is not $\mathrm{SL}(2,\RR)$ with a left-invariant metric. In this case, the orbit $X$ of $\mathrm{SL}(2,\RR)$ is a two-dimensional homogeneous space, either the two-dimension\-al de Sitter 
dS$_2$
or the Poincar\'e disk 
H$_2$. We will not make a distinction between the two-dimensional de Sitter and the two-dimensional anti-de Sitter. The deformed space is at that value of $t$ isometric to $X\times \RR$ and its isometries are precisely ${\mathfrak{sl}_2}\oplus\RR$. The space is homogeneous but with a one-dimensional isotropy.

But how can the cosets $X\times \RR$ be realized as deformations of anti-de Sitter? Let us consider a family of spacetimes $(M_t,g_t)$ where at $t=0$ we place anti-de Sitter and at $t=1$ we place $X\times \RR$, with $X=\mathrm{AdS}_2$ or $H_2$. We can either take the spaces $(M_t,g_t)$ for $0<t< 1$ as isometric or not. We will call the limit $t\rightarrow 0$ a spacetime limit of the family. If the spaces for $0<t<1$ are isometric, we call it a limit of a fixed spacetime. Geroch gives various hereditary properties for spacetime limits~\cite{Geroch:1969ca}. One of them is that the kernel of a conncetion $D_t$ defined on a vector bundle over $(M_t,g_t)$ cannot reduce its dimension at a spacetime limit. The theorem is especially powerful because it is based on the smoothness of the holonomy of $D_t$ rather than the continuity of the kernel. We can then show that anti-de Sitter and $X\times\RR$ are not infinitesimally connected.

Since $X\times \RR$ has a parallel vector, the connection being the Levi-Civita, anti-de Sitter space cannot be a spacetime limit of $X\times \RR$. Since the isometry algebra cannot reduce its dimension, the connection being the Killing transport connection on $\Lambda T^* M\oplus\Lambda^2 T^* M$, $X\times \RR$ cannot be a 
spacetime limit of anti-de Sitter either. We therefore need a non-isometric 
family of homogeneous spaces $(M_t,g_t)$ for $0<t<1$ ``in between'' the two homogeneous spaces. We have shown that all infinitesimal deformations of anti-de Sitter are given by left-invariant metrics on $\mathrm{SL}(2,\RR)$ or $\mathrm{\mysocapital}(1,1;t)$.

\subsection{The sphere}
The case of the sphere is slightly simpler because all one-parameter subgroups of $\mathrm{SU}(2)$ are conjugate to each other. The isometry algebra of the sphere is $\mathfrak{so}(4)=\mathfrak{su}_2\oplus\mathfrak{su}_2$ and its subalgebras of dimension at least three are isomorphic to either $\mathfrak{su}_2\oplus\mathfrak{su}_2$, $\mathfrak{su}_2\oplus\RR$ or $\mathfrak{su}_2$. These are all rigid under deformations. If $\mathfrak{su}_2$ in $\mathfrak{su}_2\oplus\RR$ does not act transitively on the deformation, the space is locally $S^2\times \RR$. This latter space can be achieved as a finite deformation. Similarly to what was described in section \ref{sec:so22}, infinitesimal homogeneous deformations of the sphere can only be given by $\mathrm{SU}(2)$ with a left-invariant metric. 

\subsection{de Sitter and hyperbolic space}
Hyperbolic space, $\mathrm{SL}(2,\CC)/\mathrm{SU}(2)$, and de Sitter space, $\mathrm{SL}(2,\CC)/\mathrm{SL}(2,\RR)$, share the same isometry algebra of $\mathfrak{sl}(2,\CC)$. We will be signature agnostic and study homogeneous deformations for both spaces simultaneously. In fact, we show that the two spaces are homogeneously rigid in TMG.

The subalgebras of $\mathfrak{sl}(2,\CC)$ up to conjugation are known, see for instance \cite{Patera:1974zd}. The subalgebras of dimension at least three are up to conjugation: the four-dimensional Borel subalgebra $\mathfrak{b}_4$, the three-dimensional $\mathfrak{sl}_2$ and $\mathfrak{su}_2$, and the one parameter family of three-dimensional algebras $\mathfrak{\mysosmall}(2;\theta)$, $\theta\in[0,\frac{\pi}2]$. Let us define the matrices that make up the Borel subalgebra $\mathfrak{b}_4$  of $\mathfrak{sl}(2,\CC)$:
\begin{align*}
 l_1 &= \begin{pmatrix} 1 &0\\0&-1 \end{pmatrix}
,&
l_2 &= \begin{pmatrix} i &0\\0&-i \end{pmatrix},&
m_1 &= \begin{pmatrix} 0 &1\\0&0 \end{pmatrix},
&
m_2 &= \begin{pmatrix} 0 &i\\0&0 \end{pmatrix}~.
\end{align*}
Then, $\mathfrak{\mysosmall}(2;\theta)$ is spanned by $l=\cos\theta\, l_1+\sin\theta \,l_2$, and the $m_i$. The subalgebras $\mathfrak{\mysosmall}(2;0)$, $\mathfrak{su}_2$ and ${\mathfrak{sl}_2}$ are not transitive on de Sitter. Similarly, the subalgebras $\mathfrak{\mysosmall}(2;\frac\pi2)$, ${\mathfrak{sl}_2}$ and $\mathfrak{su}_2$ are not transitive on hyperbolic space.

We show in the appendix that $H^2(\mathfrak{b}_4;\mathfrak{b}_4)=0$, and so $\mathfrak{b}_4$ is rigid under deformations. Apriori, a possibly non-trivial homogeneous deformation $(B_4/H_t,B)$ is given by deforming the one-dimensional $H_t\subset B_4$ and the $\mathfrak{h}_t$-invariant metric $B_t$ on $\mathfrak{b}_4/\mathfrak{h}_t$. A one-dimensional subalgebra $\mathfrak{h}\subset\mathfrak{b}_4$ is up to conjugacy spanned by either  $\cos\theta' \, l_1+\sin\theta' \, l_2 $ or $m_1$. The element $m_1$ acts on $\mathfrak{b}_4/\langle{ m_1 } \rangle$ as 
\begin{align}
  [m_1,l_1] & = -2m_1 \equiv 0 \mod m_1,\\
  [m_1,l_2] & = -2 m_2~,\\
  [m_1,m_2] & =  0
~.
\end{align}
It follows that a symmetric bilinear form on $ \mathfrak{b}_4/\langle{ m_1 } \rangle$ is necessarily degenerate and thus $\mathfrak{b}_4/\langle{ m_1 } \rangle$ does not admit an $m_1$-invariant metric. In a similar manner we can show that, unless $\mathfrak{h}$ is spanned precisely by $l_2$, there can be no $\mathfrak{h}$-invariant metric on $\mathfrak{b}_4/\mathfrak{h}$. 

When $\mathfrak{h}$ is spanned by $l_2$, there is a reductive split $\mathfrak{b}_4/\mathfrak{h}=  \langle  l_1 \rangle \oplus \langle m_1 , m_2 \rangle $. Recall that $l_2$ acts on the $m_i$ as an infinitesimal rotation in $\mathfrak{so}(2)$. The most general $l_2$-invariant metric on $\mathfrak{b}_4/\mathfrak{h}$ 
is thus given up to rescalings by
\begin{equation}\label{eq:BonB4modH} B^{-1} =+ m_1\,m_1+m_2\,m_2 +\frac{1}{\mu} l_1\, l_1  ~.~\end{equation}
However, the metric \eqref{eq:BonB4modH} is precisely de Sitter or hyperbolic space for respectively $\mu<0$ or $\mu>0$. We show this explicitly in the appendix \ref{app:borel}. The problem now reduces to the deformations of $\mathfrak{\mysosmall}(2;\theta)$. But as we have already seen, there are no anisotropic deformations of de Sitter or hyperbolic space that continue to solve TMG.

\subsection{Coset limits}\label{sec:cosets}
We gave arguments that the infinitesimal homogeneous deformations of the maximally symmetric spaces solve TMG only within the class of anisotropic homogeneous deformations. However, one can take a finite deformation where the space is driven to $X\times\RR$, with $X$ one of $\mathrm{AdS}_2$, $H_2$ or $S^2$. These finite deformations are always possible by taking an extremal limit of the type D spacetimes. 

In order to ``break free'' from the confines of the anisotropic solutions and their infinitesimal deformations, one should take a limit where the metric $B$ on the three-dimensional Lie algebra becomes degenerate. We may call such a limit in this context an extremal limit. For instance, if we replace in the metric \eqref{eq:warpeds} of spacelike warped anti-de Sitter $u'=\sqrt{|\gamma|}\,u$,
\begin{equation}\label{eq:warpedspre}
 g = \alpha\left( -\cosh^2\sigma\grad t^2 + \grad\sigma^2\right)\pm\left(\grad u'+ \sqrt{|\gamma|} \cosh\sigma\grad t\right)^2~,
\end{equation}
and send $\gamma \rightarrow 0$, then we arrive at $\mathrm{AdS}_2\times \RR$. At $\gamma=0$ the isometry is still ${\mathfrak{sl}_2}\oplus\RR$ but ${\mathfrak{sl}_2}$ does not act transitively. Similarly, one can obtain $\mathrm{H}_2\times\RR$ and $\mathrm{S}^2\times\RR$ of either signature from the timelike warped anti-de Sitter and the stretched / squashed sphere, respectively. These spaces are non-Einstein and conformally flat. That is, they solve the pure Cotton equation $C_{ab}=0$, or $\lim_{\gamma\rightarrow0}\mu=\infty$ in the type D solutions, see for instance \eqref{eq:RmuSWAdS}.

More generally, at any limit of a family of spacetimes the isometry group continues to act as a group of isometries. The isometries cannot reduce in dimension although they can potentially contract\footnote{an extremal limit where the Lie algebra contracts but continues to act transitively is given by setting  in \eqref{eq:warpedspre} $\alpha=1/\epsilon^2$, $\gamma=\gamma'/\epsilon^2$, $\sigma'=\epsilon\sigma$ and $t'=\epsilon t$ and sending $\epsilon\rightarrow0$. We obtain the warped flat space as a limit in which ${\mathfrak{sl}_2}$ contracts to $\mathfrak{so}(1,1)\ltimes\RR^{1,1}=\mathfrak{\mysosmall}(1,1;\frac\pi2)$.}. One might wonder if similar limits are obtained for the other anisotropic solutions. That is, whether there is an extremal limit in which the three-dimensional Lie algebra ceases to act transitively. A maximally symmetric two-dimensional metric can be invariant at most under ${\mathfrak{sl}_2}$, $\mathfrak{su}_2$, $\mathfrak{\mysosmall}(1,1;\frac\pi2)=\mathfrak{so}(1,1)\ltimes\RR^{1,1}$, 
or $\mathfrak{\mysosmall}(2;\frac\pi2)=\mathfrak{so}(2)\ltimes\RR^2$. The three-dimensional metric is always of the form $f(x) \grad s^2_X + \grad x^2$ where $X$ is one of $\mathrm{AdS}_2$, $H_2$, $S^2$, $\RR^{2}$ or $\RR^{1,1}$. The space is conformally flat, at most Einstein. 

All infinitsimal homogeneous deformations of anisotropic solutions are within the class of anisotropic solutions. This follows from the infinitesimal deformations of the triples $(G,H,B)$. Warped anti-de Sitter and the stretched / squashed sphere have enhanced isometry $\mathrm{SL}(2,\RR)\times\RR$ and $\mathrm{SU}(2)\times\RR$. The simple Lie algebra will cease to act transitively only at some finite value of the deformation parameter. In this case we obtain a solution of the form $X\times\RR$ that is conformally flat. All other deformations are anisotropic. 

\section{Discussion}\label{sec:discussion}

In this first part of this paper we solved for the homogeneous anisotropic solutions of topologically massive gravity with a potentially non-vanishing cosmological constant. To our knowledge this has not been presented before. However, we do not find any new solutions than the ones that have been found with a variety of other means. The novelty of section \ref{sec:3dGeometry} is that we solved this problem systematically. We also studied their homogeneous deformations. We showed that
\begin{itemize}
 \item infinitesimal homogeneous deformations of homogeneous anisotropic solutions are anisotropic,
 \item out of the maximally symmetric spaces only anti-de Sitter, the sphere, and flat space can be infinitesimally homogeneously deformed.
\end{itemize}
A motivation for this would be to place the deformations of the maximaly symmetric spaces in a potentially wider context. For instance, how can one deform the holography of anti-de Sitter by retaining as many symmetries as possible.

The methodology that we used for the first part differs from that of~\cite{Ortiz:1989vc}. Ortiz in \cite{Ortiz:1989vc} fixes an orthonormal frame up to local Lorentz rotations in  $\mathrm{SO}(1,2)$.  Let the structure coefficients in a given orthonormal basis $e_i$ be $c_{ij}{}^k=\left( b^{(kl)}+ \epsilon^{klm} a_m\right) \epsilon_{ijl}$. The Jacobi identity is simply $b^{ij}a_j=0$. The action of the orthogonal group results to different types (classes) for the space of orbits for the $b^{ij}$ and $a^i$. The equations of motion of TMG reduce to algebraic equations on the structure coefficients of the frame, where the types for the $b^{ij}$ and $a^i$ have been fixed. One works similarly in euclidean signature with $\mathrm{SO}(1,2)$ replaced by $\mathrm{SO}(3)$. The riemannian geometry of three-dimensional groups has been studied in this way in the classic work of Milnor~\cite{Milnor:76}, and more recently three-dimensional pseudo-riemannian homogeneous geometries were classified similarly in~%
\cite{Calvaruso:07,*Calvarusoadd:08}. %
When the Lie algebra is $\mathfrak{g}={{\mathfrak{sl}_2}}=\mathfrak{so}(1,2)$ and for lorentzian signature, or $\mathfrak{g}=\mathfrak{su}_2$ for euclidean signature, both $B_{ab}$ and $b_{ij}$ are diagonalizable to $\eta_{ij}$, or respectively $\delta_{ij}$, via a $\mathrm{SO}(3)$ rotation but not simultaneously.

We did not attempt to make contact with Ortiz's methodology, because the change of basis would not be straightforward. Fixing an orthonormal basis as in \cite{Ortiz:1989vc} is a preferred method for reasons of generality: our method appears to be taxonomically lengthier. On the other hand there are some obvious benefits in our method. Fixing the Lie algebra basis for a given Lie algebra is immediate in writing down coordinate forms for the metric. 
Finally, we can keep track of each Lie algebra seperately, which is useful in the second part of the paper. In fact, our project began with metrics on $\mathrm{\mysocapital}(2;\theta)$, the family of transitive three-dimensional isometries of de Sitter and hyperbolic space. 

\begin{figure}[ht]
\begin{center}
\subfloat[zoomed out]{\includegraphics[
]{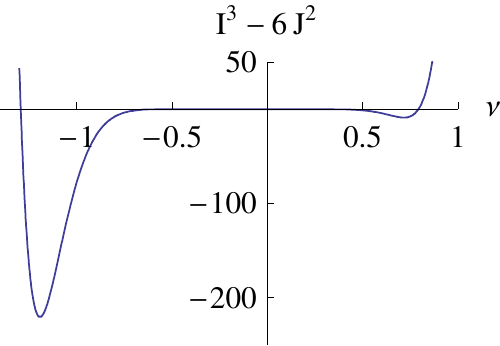}\label{fig:IJ1}}
\quad
\subfloat[second root]{\includegraphics[]{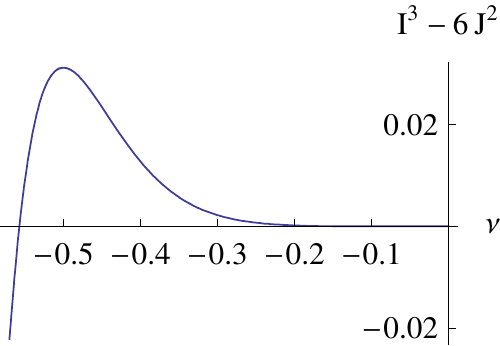}\label{fig:IJ2}}
\quad
\subfloat[fourth root]{\includegraphics[]{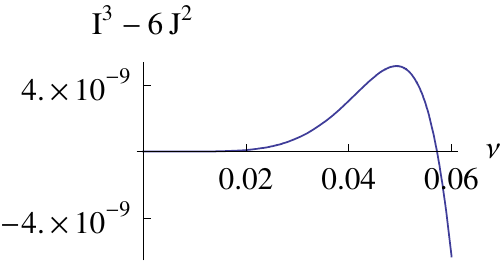}\label{fig:IJ3}}
\caption{The graph of $I^3-6 J^2$ for \eqref{eq:bsyzygies} zoomed in different regions.}\label{fig:IJ}
\end{center}
\end{figure}

It was stated in \cite{Chow:2009km}, but also in \cite{Bakas:2010kc}, that no known solution exists that is of type I$_\CC$, or equivalently that Ortiz's (b) solution in \cite{Ortiz:1989vc} is only of type I$_\RR$. In our work the solution \ref{item:sl2typeIC} is precisely of type I$_\CC$. Although we have not attempted to translate our solutions to Ortiz's formulation, we take verbatim the curvature invariants for $\mu=1$ of his (b) solution
\begin{subequations}\begin{align}
 I&=R_a^b R_b^a = 8 \nu^4 + 2 \nu^2 + 4 \nu^3~,\\
 J&=R_a^b R_b^c R_c^a = 12 \nu^5 + 6 \nu^4 + \nu^3~.
\end{align}\label{eq:bsyzygies}\end{subequations}
There are some conditions on the range of the parameter $\nu$ so that this solution is genuinely different to others, but we need not worry about this here and take $\nu\in\RR$. As explained in \cite{Chow:2009km}, types O, N and III satisfy $I=J=0$, types D and II satisfy $I^3=6 J^2\neq0$, and types I$_\RR$ and I$_\CC$ satisfy respectively $I^3>6 J^2$ and $I^3<6 J^2$. For \eqref{eq:bsyzygies}, $I^3-6 J^2$ has two negative roots and two positive roots, see figure~\ref{fig:IJ}, and its sign alternates between these roots. 

Ortiz finds a one-parameter family on ${{\mathfrak{sl}_2}}$ in his solution (a) (case $a=0$), which can be manifestly written in the triaxial form \ref{item:sl2triaxial}, a one-parameter family on ${{\mathfrak{sl}_2}}$ of type I in his solution (b), a solution on $\mathfrak{sl}_2$ of type I$_\RR$ in the first solution in (c), and a one-paramater family on $\mathfrak{\mysosmall}(1,1;\theta)$ of type N (or apriori type III) in his second solution in (c). According to our result we should identify his second solution in (c) with the flat space pp-wave. We identify solutions purely due to their spacetime type and symmetries: solutions of type I are identified with \ref{item:sl2triaxial} and \ref{item:sl2typeIC}, or the unified family in \eqref{eq:R0unified}.

The equations of generalized massive gravity (GMG) are the addition of a traceless quadradic curvature form to the Einstein-Cotton equation that is derived from an action. The authors of \cite{Bakas:2010kc} solve the equations of GMG on $\mathrm{SU}(2)$. They also analytically continue to metrics on $\mathrm{SL}(2,\RR)$. In both cases, they compare with the known solutions of TMG by turning off the quadratic term. In this way, the stretched / squashed sphere and triaxially deformed sphere were continued to the spacelike and timelike warped anti-de Sitter and the triaxially deformed anti-de Sitter. The metrics on $\mathrm{SL}(2,\RR)$ such that $\eta^{ac}B_{cb}$ has Segre type $\{12\}$, $\{1z\bar{z}\}$ or $\{3\}$ were not obtained. The continuation $\mathfrak{su}_2\rightarrow\mathfrak{sl}_2$ is a Cayley transformation $\tau_3=i\,\tau_0$, so one would have to analytically continue the parameters $\alpha$, $\beta$, $\gamma$ to complex values in order to obtain all metrics on $\mathrm{SL}(2,\RR)$. Here, we have 
solved directly on $\mathrm{SL}(2,\RR)$.

In regards to generalizing TMG to higher curvature, we only briefly looked into metrics on $\mathrm{SL}(2,\RR)$ that satisfy an equation of the form
\begin{equation}\label{eq:higher}
R_{ab} + \Lambda\,g_{ab} + \mu_1\,C_{ab} + \mu_2\, R_a^c\,R_{cb} = 0~,
\end{equation}
but we did not report on it here. We found that any metric $B_{ab}$ on ${{\mathfrak{sl}_2}}$ solves the theory, in which case $\mu_1$ and $\mu_2$ are algebraically determined from $B_{ab}$ and $\Lambda=0$ always. The generalization \eqref{eq:higher} is not derived from a lagrangian, since derivatives on curvature tensors are not included. Although \eqref{eq:higher} can be interesting in itself, a difficulty we encountered was that inverting the map $B\mapsto (\mu_1,\mu_2)$ is not immediate and in general the map is not surjective on all the real numbers $\mu_1$, $\mu_2$.

In this paper we have discussed the homogeneous deformations of the anisotropic solutions at some length. One of our results is that de Sitter and hyperbolic space cannot be homogeneously deformed in TMG. In \cite{Anninos:2011vd} equation (2.4), the metric of spacelike warped anti-de Sitter is written as
\begin{equation}\label{eq:warpeddS}
g = \frac{\ell^2}{3-\nu^2}\left( - \frac{\grad t^2}{1+t^2} + (1+t^2)\grad\phi^2+\frac{4\nu^2}{3-\nu^2}\left(\grad u + t\,\grad\phi\right)^2\right)~.
\end{equation}
It corresponds to our parameters $\beta=-\alpha=\ell^2/(3-\nu^2)$ and $\gamma=\ell^2 4\nu^2/(3-\nu^2)^2$ in \eqref{eq:warpeds} and \eqref{eq:segre111}. Note that the authors of \cite{Anninos:2011vd} call the metric \eqref{eq:warpeddS} as that of ``warped de Sitter'' when $\nu^2<3$ ($\alpha>0$) because of its properties. We have only defined warped anti-de Sitter in section \ref{sec:known}, and our definition is in clash with that of \cite{Anninos:2011vd}.

Perhaps our choice of nomenclature is confusing, since a warped metric usually refers to a fibered metric. Spacelike warped anti-de Sitter is a fibered metric over two-dimensional anti-de Sitter, which is also two-dimensional de Sitter if we swap the signature of the base metric from $(+,-)$ to $(-,+)$. Our definition is based on the isometries of $\mathfrak{sl}_2$ and spacelike warped anti-de Sitter, the metric \eqref{eq:warpeddS} for all $\nu$, has isometries $\mathfrak{sl}_2\oplus\RR$ with the $\RR$ generated by a hyperbolic element of $\mathfrak{sl}_2$ acting on the right of $\mathrm{SL}(2,\RR)$ equipped with a left-invariant metric. At the same time, spacelike (and timelike or null) warped anti-de Sitter turns out to be a deformation of anti-de Sitter, see the discussion below \eqref{eq:warpedd}. In the first version of the preprint in \cite{Anninos:2011vd} it was written (in a typographical slip) that \eqref{eq:warpeddS} reduces to de Sitter at $\nu=1$, which is wrong: it is easy to 
show 
that \eqref{eq:warpeddS} is not Einstein unless $\nu^2=-1$, in which case it is anti-de Sitter of mostly minus signature. It cannot be a deformation of de Sitter because de Sitter is rigid under deformations in TMG. 

In \cite{Anninos:2009jt} the metric
\begin{equation}\label{eq:warpedH}
g = \frac{\ell^2}{3-\nu^2}\left( \grad\rho^2 + \cosh^2\rho \grad\tau^2+\frac{4\nu^2}{3-\nu^2}\left(\grad u -\sinh\rho \,\grad\tau\right)^2\right)~ 
\end{equation}
is called a squashed three-dimensional hyberbolic geometry. It is a fibration over two-dimensional hyperbolic space, whose isometries preserve the field-strength of the Kaluza-Klein gauge field $\grad (\sinh\rho \,\grad\tau) = \dvol_{H_2}$. Along with the translation along the fiber $\partial_u$, the isometries of the three-space make up $\mathfrak{sl}_2\oplus\RR$. The metric \eqref{eq:warpedH} is none other than what we call timelike warped anti-de Sitter with $\beta=\gamma=\ell^2/(3-\nu^2)$ and $\alpha=\ell^2 4\nu^2/(3-\nu^2)^2>0$. It cannot be a deformation of three-dimensional hyperbolic geometry because the latter is rigid under homogeneous deformations in TMG.

A natural question then arises. What happens if we try to deform de Sitter or hyperbolic space in a manner as closely as possible to what is done with warped anti-de Sitter or the squashed/stretched sphere. Let us take for definiteness de Sitter with metric $g_{dS}$ and choose a Killing vector $\xi\in\mathfrak{sl}(2,\CC)$. We calculate the metric dual $\theta= g_{dS}(-,\xi)$ and investigate the deformed metric
\begin{equation}\label{eq:defdS}
g = g_{dS} + \lambda \,\theta\otimes \theta~.
\end{equation}
We can immediately say that the isometries of $g$ are given \emph{at least} by the Killing vectors of the undeformed space that commute with $\xi$: $\xi$ and $J\,\xi$ where $J$ is the complex unit in $\mathfrak{sl}(2,\CC)$. The isometries of $g$ are at least $\RR\oplus\RR$. At first sight this is less than warped anti-de Sitter, where the centralizer of $\xi$ was $\mathfrak{sl}_2\oplus\RR$. However we can do more. The algebra  $\mathfrak{sl}(2,\CC)$ has two conjugacy classes: a complex line and a fixed element. We therefore act on the space with a diffeomorphism that is the exponential of $\mathfrak{sl}(2,\CC)$ such that we bring $\xi$ to the form of a given Killing vector representative. If the de Sitter metric (we fix $R=6$) is
\begin{equation}
 g_{dS} = \frac{\grad b_1^2+\grad b_2^2-\grad z^2}{z^2}=\frac{\grad r^2+r^2\,\grad \phi^2-\grad z^2}{z^2}~,
\end{equation}
a Killing vector can be brought to one of the two forms
\begin{align}
 \xi_1 &= a( r\,\partial_r + z\,\partial_z)+ b\,\partial_\phi \\
 \xi_2 &= \partial_{b_1}~.
\end{align}
If we plug \eqref{eq:defdS} with these two choices into TMG, we find a solution only for $\xi_1$, $\lambda\, b^2=-\frac{15}{16}$ and $\lambda\, a^2=-\frac{1}{12}$. Unlike warped anti-de Sitter, this metric is not a continuous deformation.

Let us make here a final comment on warped anti-de Sitter. A global parametrization of $\mathrm{SL}(2,\RR)$ is given by the asymmetric coordinates of \cite{Coussaert:1994tu} that give the spacelike warped anti-de Sitter metric that we use in \eqref{eq:warpeds}. On the other hand, accelerating observers are described\footnote{
The global coordinates are related to the representative $\mathcal{V}=e^{{\tau}\tau_0}e^{{\sigma} \tau_1}e^{{u}\tau_2}$ and the accelerating to $\mathcal{V} = e^{\tau \, \tau_2} e^{\sigma \, \tau_1} e^{u \, \tau_2}$, see also  \cite{Jugeau:2010nq} for a discussion on the different coordinate systems.} by
\begin{equation}\label{eq:accel}
g = \beta\left(\frac{\grad x^2}{x^2-b^2} - (x^2-b^2)\grad t^2\right)+\gamma\left(\grad u + x\,\grad t\right)^2~.
\end{equation}
The constant $b\neq0$ is unphysical and can be gauged away to $b=1$. It measures the acceleration of the observers in this coordinate system. Quotienting these metrics by a finite isometry generated by $\partial_u$ one obtains the self-dual quotient. Let us remark that the self-dual quotients of spacelike warped anti-de Sitter (for negative or positive $\alpha$) appear at fixed polar angle in the near-extremal near-horizon limits of the four-dimensional Kerr black hole (for negative or positive cosmological constant), see for instance \cite{Anninos:2010gh}. But also timelike warped anti-de Sitter appears likewise at a ``polar limit'' of the Kerr black hole with NUT parameter~\cite{Mitsuka:2011bf}. In these limits one always lands with accelerating coordinates, which characterizes the original black hole horizon. But at the same time one has ``washed away'' infinity. That is, in these limits there is still no way of producing a meaningful acceleration $b$. This comment is in contradiction to some other works 
on the self-dual quotient.

\section*{Acknowledgements}
The author would like to acknowledge Patricia Ritter's contribution to sections \ref{sec:so2theta} and appendix \ref{app:borel}. He also thanks the anonymous referee who pointed out an error in the deformations into $\mathfrak{su}_2$. The author acknowledges financial support from National Chiao-Tung University, Taiwan, where the bulk of the research was done. The final revision was done under a Riemann Fellowship, ITP, University of Hannover. 

\appendix
\section{Subgroups of \texorpdfstring{$\mathrm{SL}(2,\RR)\times\mathrm{SL}(2,\RR)$}{{SL}(2)x{SL}(2)}}\label{app:goursatcosets}
We will use the ${{\mathfrak{sl}_2}}$ basis $\tau_a$, $[\tau_a,\tau_b]=\epsilon_{ab}{}^c\tau_c$,  that induce the left (respectively right) action on $\mathrm{SL}(2,\RR)$ with vector fields $l_a$ (respectively $r_a$). We want to find all subgroups $G_0$ of $\mathrm{SL}(2,\RR)_L\times \mathrm{SL}(2,R)_R$ of dimension at least three, and in particular those that are transitive on $\mathrm{SL}(2,\RR)$. 

Goursat's lemma gives the subgroups $G_0$ of a direct product $G_1\times G_2$. Assume that $G_i = \proj_i G_0$ under $G_0\subseteq G_1\times G_2$ and let $N_2=\ker \proj_1$, $N_1=\ker\proj_2$. It can be shown that $N_i$ is a normal subgroup of $G_i$. The lemma asserts that the image of ${G_0}{}$ in $G_1/N_1\times G_2/N_2$ is the graph of an isomorphism $G_1/N_1\approx G_2\times N_2$.

We use the lemma by taking each $G_i$, $i=1,2$, as one of the subgroups of $\mathrm{SL}(2,\RR)$: $\mathrm{SL}(2,\RR)$, $\RR^+\ltimes\RR$, $\RR_x$ and $1$. If $G_i=\mathrm{SL}(2,\RR)$, being simple, $N_i=\mathrm{SL}(2,\RR)$ or $1$. The only normal subgroup of $\RR^+\ltimes\RR$ is $\RR$, so if $G_i=\RR^+\ltimes\RR$ then $N_i=1,$ $\RR$ or $\RR^+\ltimes\RR$. If $G_i=\RR_x$ then $N_i=\RR_x$ or $1$. We then scan for the cases in which $G_1/N_1\approx G_2/ N_2$, where we also require $\dim {G_0}{} = \dim G_1 + \dim N_2 \geq 3$.

Say $G_1\times G_2 = \mathrm{SL}(2,\RR)\times\mathrm{SL}(2,\RR)$. Either $N_i=1$ and 
\[ {G_0}{}= \left\{ (g,h g h^{-1}) 
\text{ for } g\in\mathrm{SL}(2,\RR)\right\}\approx \mathrm{SL}(2,\RR), \]
where $h\in\mathrm{SL}(2,\RR)$ is an automorphism $\mathrm{SL}(2,\RR)_L\approx\mathrm{SL}(2,\RR)_R$, or $N_i=\mathrm{SL}(2,\RR)$ and 
${G_0}{}=\mathrm{SL}(2,\RR)\times\mathrm{SL}(2,\RR)$. Say $G_1\times G_2 = \mathrm{SL}(2,\RR)\times \left(\RR^+\ltimes\RR\right)$. The only possibility is $N_i=G_i$ and ${G_0}{}=\mathrm{SL}(2,\RR)\times \left(\RR^+\ltimes\RR\right)$. If $G_i\times G_2 =\mathrm{SL}(2,\RR)\times \RR_x$ then $\dim N_1 \geq 1$ requires $N_1 = \mathrm{SL}(2,\RR)$ and ${G_0}{} =\mathrm{SL}(2,\RR)\times \RR_x$. If $G_2=1$ then $H = \mathrm{SL}(2,\RR)\times 1$.

If $G_1\times G_2 =\left(\RR^+\ltimes\RR\right)_L \times \left(\RR^+\ltimes\RR\right)_R$, we can have two cases on dimensional grounds. In the first case, $N_1=G_1$ and ${G_0}{}=\left(\RR^+\ltimes\RR\right)_L \times \left(\RR^+\ltimes\RR\right)_R$. In the second case, $N_1=\RR$ and the isomorphism $(\RR^+)_L \approx (\RR^+)_R$ is given by a non-zero real number $\mu$. Its Lie algebra is precisely
\begin{align}
{\mathfrak{g}_0} &= \left\langle  l_2+\mu r_2, \, r_0+ r_1, \, l_0+l_1 \right\rangle \approx \mathrm{\mysocapital}(1,1;\theta)~,\\
\intertext{ with }
\theta&=\arctan\left|\frac{1-\mu}{1+\mu}\right|\neq \frac\pi4 . 
\end{align}
If $G_1\times G_2 = \left(\RR^+\ltimes\RR\right)_L \times
\RR_x$, then for dimensional reasons 
\begin{equation}
 {G_0}{} =  \left(\RR^+\ltimes\RR\right)_L \times
\RR_x\approx\mathrm{\mysocapital}(1,1;\theta=\frac{\pi}{4})~.
\end{equation}

The subgroups of dimension at least three that appear are: $\mathrm{SL}(2,\RR)\times\mathrm{SL}(2,\RR)$, $\mathrm{SL}(2,\RR)\times \left(\RR^+\ltimes\RR \right) $, $\mathrm{SL}(2,\RR)\times \RR_x$, $\left(\RR^+\ltimes\RR \right)\times \left(\RR^+\ltimes\RR \right)$,  $\mathrm{SL}(2,\RR)$, 
and $\mathrm{\mysocapital}(1,1;\theta)$. Note that we simply list the groups that appear, rather than the distinct subgroups under conjugacy. In order to identify which subgroups act transitively on anti de Sitter, we simply identify $r_a \equiv l_a$ at the identity of $\mathrm{SL}(2,\RR)$, and require that the image of $\mathfrak{g}_0$ be larger or equal to three. The subgroups of $\mathrm{SL}(2,\RR)_L\times\mathrm{SL}(2,\RR)_R$ that act transitively on anti de Sitter are $\mathrm{SL}(2,\RR)$ and $\mathrm{\mysocapital}(1,1;\theta)$ with $\theta\neq \frac\pi2$.

\section{Metrics on \texorpdfstring{$B_4/H$}{B4/H}}\label{app:borel}
Let us begin by finding the conjugacy classes of one-dimensional subgroups of $B_4$. The Borel subgroup $B_4$ of $\mathrm{SL}(2,\CC)$ is 
\begin{equation} B_4 = \left\{ \begin{pmatrix} s & b\\0&1/s \end{pmatrix}\text{ with } s,b\in\CC \text{ and } s\neq 0\right\} ~.\end{equation}
Let us write the elements in the Lie algebra $\mathfrak{b}_4$
\[\begin{pmatrix} a & b\\0&-a \end{pmatrix}~, \]
with $a,b\in\CC$, in the shorthand $(a,b)$. 

The adjoint action on the Lie algebra is
\begin{equation} \begin{pmatrix}
    s' & b' \\ 0& 1/s'
   \end{pmatrix}(a , b )\begin{pmatrix}
    1/s' & -b' \\ 0& s'
   \end{pmatrix} =( a , s'^2 b -2a \,s'\,b' )~.
\end{equation}
If $a\neq 0$, we can choose $b'=(s'^2 b-1)/(2 a s')$ so that $(a,b) \stackrel{\text{cong.}}{=} (a,1)$. If $a =  0$, we choose $s'\neq 1/\sqrt{b}$ so that again $(0,b) \stackrel{\text{cong.}}{=} (0,1)$. That is, the conjugacy classes $(a,b)\stackrel{\text{cong.}}{=} (a,1)$ are parametrized by $a\in\CC$. However, we shall do a bit more. The generator of a one-dimension subgroup $(a,1)$ is defined up to multiplication by a real non-zero number, which allows us to fix $|a|=1$ or else $a=0$. Furthermore, if $a\neq 0$ then $(a,1) \stackrel{\text{cong.}}{=} (a,0)$, by choosing e.g. $b'=1/(2a)$ and $s'=1$ above. The conjugacy classes we consider are $(a,0)$ with $|a|=1$ and $(0,1)$. 

Only the class $(i,0)=l_2$ is such that, with $\mathfrak{h}$ spanned by $l_2$, the space $\mathfrak{b}_4/\mathfrak{h}$ allows a non-degenerate metric. A reductive split is $\mathfrak{b}_4=\mathfrak{k}\oplus\mathfrak{h}$ with
\[ \mathfrak{k} = \langle l_1,m_1,m_2 \rangle .\]
It is an easy exercise to show that the most general non-degenerate $\mathfrak{h}$-invariant metric $B$ on $\mathfrak{k}$ is up to rescalings given by
\begin{equation} B =+ \tilde{m}_1\,\tilde{m}_1+\tilde{m}_2\,\tilde{m}_2 +{\mu}\, \tilde{l}_1\, \tilde{l}_1  ~,\end{equation}
where we have used the dual basis $\{\tilde{m}_1,\tilde{m_2},\tilde{l}_1\}$ of $\mathfrak{k}^*$. The element $l_2$ acts as euclidean rotations on the two-dimensional subspace $\langle \tilde{m}_1,\tilde{m}_2\rangle$.

We choose the coset representative
\begin{equation}\label{eq:nillleftV} \mathcal{V} =  \begin{pmatrix} 1 &b\\0& 1 \end{pmatrix}  \begin{pmatrix} R & 0\\0& 1/R \end{pmatrix} = \begin{pmatrix} R &b/R\\0& 1/R \end{pmatrix}, \end{equation}
with $b\in\CC$ and $R\in\RR\backslash\{0\}$. We calculate
\[ \mathcal{V}^{-1} \grad \mathcal{V} = \frac{\grad R}{R} l_1+ \frac{\grad b_1}{R^2}  m_1 + \frac{\grad b_2}{R^2} m_2, \]
so that the metric $g=B(\mathcal{V}^{-1} \grad \mathcal{V},\mathcal{V}^{-1} \grad \mathcal{V})$ is given by
\begin{equation}\label{eq:nillleftmetric} g= +\frac{\grad b_1^2}{R^4}    +\frac{\grad b_2^2}{R^4}\pm |\mu| \frac{\grad R^2}{R^2}.\end{equation}
By setting $z= |\mu|^{-1/2}R^2$ we identify
\[ g= |\mu|\frac{\grad b_1^2+\grad b_2^2 \pm \grad z^2}{z^2} \]
as hyperbolic or de Sitter space in familiar Poincar\'e coordinates.

\section{Lie algebra cohomology}\label{app:coho}

Infinitesimal deformations of a Lie algebra $\mathfrak{g}$ are interpeted in terms of the cohomology group $H^2(\mathfrak{g},\mathfrak{g})$. Let us briefly note some other useful descriptions of a Lie algebra that are encoded by  cohomology. The outer automorphisms of $\mathfrak{g}$ are elements of $H^1(\mathfrak{g},\mathfrak{g})$. The central extensions of $\mathfrak{g}$ make up $H^2(\mathfrak{g},\RR)$ and $H^1(\mathfrak{g},\RR)$ is isomorphic to the abelian ideal $\mathfrak{g}/[\mathfrak{g},\mathfrak{g}]$. A semisimple Lie algebra has $H^1(\mathfrak{g},\RR)=H^2(\mathfrak{g},\RR)=H^2(\mathfrak{g},\mathfrak{g})=0$. We henceforth write $H^n(\mathfrak{g}):=H^n(\mathfrak{g},\RR)$. Our aim in this section is to show that the transitive subalgebras of the anti-de Sitter isometry and the Borel subalgebra of $\mathfrak{sl}(2,\CC)$ are rigid. 

Let us calculate some cohomology groups for the non-abelian two-dimensional Lie algebra $\RR\ltimes\RR$. A basis for the Lie algebra is $x,y$ with $[x,y]=y$ and the dual basis is $\tilde{x},\tilde{y}$. The one-dimensional $C^2(\mathfrak{g},\RR)$ is spanned by $\tilde{x}\wedge\tilde{y}$ and is a cocycle, $\grad(\tilde{x}\wedge\tilde{y})=0$, on dimensional grounds. It is also a coboundary, with $\tilde{x}\wedge\tilde{y}=-\grad\tilde{y}$, as can be seen from $-\grad\tilde{y}(x,y) = \tilde{y}([x,y] )= 1$, and so $H^2(\RR\ltimes\RR)=0$. A short calculation also shows that the automorphism group is two-dimensional, and since it is centerless, $\mathfrak{aut}(\mathfrak{g})=\mathfrak{g}$ and $H^1(\RR\ltimes\RR,\RR\ltimes\RR)=0$. Being the unique two-dimensional non-abelian Lie algebra, it can deform to the abelian two-dimensional algebra, but is otherwise rigid. All together, for $\mathfrak{g}=\RR\ltimes\RR$, we have shown that
\begin{equation}\label{eq:cohoblocks}
H^2(\mathfrak{g},\mathfrak{g})=H^2(\mathfrak{g})=H^1(\mathfrak{g},\mathfrak{g})=\mathfrak{g}^\mathfrak{g}=0
\end{equation}
We have used the notation $\mathfrak{m}^\mathfrak{g}$ for the ideal in $\mathfrak{m}$ that is stable under the action of $\mathfrak{g}$. Note that the same relations \eqref{eq:cohoblocks} are still valid if we replace for the Lie algebra $\mathfrak{g}={{\mathfrak{sl}_2}}$, by virtue of it being simple. 

A homomorphism of complexes $f:C^p_1\rightarrow C^p_2$ that commutes with their differential, $\grad_2 f=f\,\grad_1$, induces a homomorphism of cohomologies $\hat{f}:H^p_1\rightarrow H_2^p$. A homomorphism $\phi:\mathfrak{m}\rightarrow\mathfrak{n}$ of $\mathfrak{g}$-modules induces a homomorphism 
${f}:C^p(\mathfrak{g},\mathfrak{m}) \rightarrow C^p(\mathfrak{g},\mathfrak{n})$ that commutes with their differentials. From this we have the isomorphism
\[ H^p(\mathfrak{g},\mathfrak{m}\oplus\mathfrak{n})\approx H^p(\mathfrak{g},\mathfrak{m})\oplus H^p(\mathfrak{g},\mathfrak{n})~.\]
In particular, a direct sum $\mathfrak{g}_1\oplus\mathfrak{g}_2$ satisfies 
\begin{equation}\label{eq:tensocoho} H(\mathfrak{g}_1\oplus\mathfrak{g}_2,\mathfrak{g}_1\oplus\mathfrak{g}_2)\approx H(\mathfrak{g}_1\oplus\mathfrak{g}_2,\mathfrak{g}_1)\oplus H(\mathfrak{g}_1\oplus\mathfrak{g}_2, \mathfrak{g}_2) .\end{equation}
This equation is the first step to show the rigidity of the direct sums in section~\ref{sec:so22}. 

Consider the two complexes $C_1=C(\mathfrak{g}_1,\mathfrak{m}_1)$ and $C_2=C(\mathfrak{g}_2,\RR)$. One has a natural filtration on $C(\mathfrak{g}_1\oplus\mathfrak{g}_2,\mathfrak{m}_1)\approx C_1\otimes C_2$ that terminates at the second page of the spectral sequence, and so we have the K\"unneth formula $H(C_1\otimes C_2)\approx H(C_1)\otimes H(C_2)$. Combining this with \eqref{eq:tensocoho} we arrive at
\begin{equation}\label{eq:defdecomp}  H^2(\mathfrak{g}_1\oplus\mathfrak{g}_2,\mathfrak{g}_1\oplus\mathfrak{g}_2) = \bigoplus_{m+n=2} \left( H^m(\mathfrak{g}_1,\mathfrak{g}_1)\otimes H^n(\mathfrak{g}_2)\oplus H^m(\mathfrak{g}_1)\otimes H^n(\mathfrak{g}_2,\mathfrak{g}_2)\right) .\end{equation}
The direct sums that appear as subalgebras of ${{\mathfrak{sl}_2}}\oplus{{\mathfrak{sl}_2}}$ and are of dimension greater than three are
\begin{equation}{{\mathfrak{sl}_2}}\oplus{{\mathfrak{sl}_2}},\, {{\mathfrak{sl}_2}}\oplus \left(\RR\ltimes\RR \right) ,\,{{\mathfrak{sl}_2}}\oplus \RR\text{ and }
\left(\RR\ltimes\RR \right)\oplus \left(\RR\ltimes\RR \right) .\end{equation}
Choosing $\mathfrak{g}_1$ and $\mathfrak{g}_2$ so that we can use \eqref{eq:cohoblocks} and \eqref{eq:defdecomp} appropriately, we have that they are all rigid. We next study the deformations of the Borel subalgebra $\mathfrak{b}_4$.

Let $\mathfrak{h}$ be a subalgebra of $\mathfrak{g}$. We define the relative cohomology $H^n(\mathfrak{g},\mathfrak{h},\mathfrak{m})$ as those cocycles $f$ in $H^n(\mathfrak{g},\mathfrak{m})$ that satisfy
\[ i_a f = a\cdot f=0\text{ for all }a\in\mathfrak{h} .\]
Let $\mathfrak{h}$ be a reductive subalgebra in $\mathfrak{g}$, the $\mathfrak{g}$-module $\mathfrak{m}$ a finite dimensional semisimple $\mathfrak{h}$-module, and the restriction homomorphism $H^p(\mathfrak{g})\rightarrow H^p(\mathfrak{h})$ be onto. The Hochshild-Serre~\cite{Hochshild-Serre} factorization theorem establishes the isomorphism
\[ H^p(\mathfrak{g},\mathfrak{m}) \approx \bigoplus_{m+n=q} H^m (\mathfrak{h},\RR)\otimes H^n(\mathfrak{g},\mathfrak{h},\mathfrak{m}) .\]
Furthermore, if $\mathfrak{g}=\mathfrak{h}\ltimes \mathfrak{i}$ and $\mathfrak{h}$ is semisimple, then one can replace $ H(\mathfrak{g},\mathfrak{h},\mathfrak{m})=H(\mathfrak{i},\mathfrak{m})^\mathfrak{g}$. 

The Borel subalgebra $\mathfrak{b}_4$ of $\mathfrak{sl}(2,\CC)$ has the reductive decomposition $\mathfrak{b}_4=\langle l_1,l_2 \rangle \ltimes \langle m_1, m_2 \rangle$. We shall use the Hochshild-Serre theorem with $\mathfrak{h}=\langle l_1,l_2 \rangle$ and $(\mathfrak{g}/\mathfrak{h})^*=\langle \tilde{m}_2,\tilde{m}_2\rangle$ in order to show that $\mathfrak{b}_4$ is rigid. Since $\mathfrak{h}$ acts effectively on $(\mathfrak{g}/\mathfrak{h})^*$, the only non-zero relative module $C^n(\mathfrak{g},\mathfrak{h},\mathfrak{m})$ is
\[ C^0( \mathfrak{g},\mathfrak{h},\mathfrak{g}) = \RR\otimes \mathfrak{h} .\]
One calculates
\[ \grad \sum_{i=1}^2 \mu_i \, 1\otimes l_i   = - \sum_{i,j=1,2}\mu_j\, \tilde{m}_i\otimes [m_i,l_j]~,\]
which is never zero, so $H^0(\mathfrak{g},\mathfrak{h},\mathfrak{m})=0$. In summary, $H(\mathfrak{b}_4,\mathfrak{b}_4)=0$ and in particular $\mathfrak{b}_4$ is rigid.

\section{Three-dimensional algebras in other notation}\label{app:groupdiff}
In table~\ref{table:morebianchi} we relate the three-dimensional algebras we use to the conventions of 
Levy-Nahas~\cite{levy-nahas:deform}, Fialowski et al.~\cite{fialowski}, Conatser~\cite{conatser:196} and Weimar-Woods~\cite{weimar-woods:2028}. 
\afterpage{%
    \clearpage
    \begin{landscape}
        \centering 
       
\begin{table}[thb]
\begin{center}
\begin{tabular}[h]{*{7}{l}}
\toprule
Bianchi      &  Levy-Nahas & Fialowski et al. & Weimar-Woods & Conatser& here& restriction \\\midrule
I  & $A_1$ & $\RR^3$ &  - &$C_1$& $\RR$&\\
II&  $A_2$ & $\mathfrak{n}_3$ & Ls &$C_2$& $\mathfrak{a}_\infty$ & \\
III =VI$_{-1}$  & $A_3=A_5(-1)$& $\mathfrak{r}_{2}\oplus\RR=\mathfrak{r}_{3,0}$ & Lssa-(-1) & $C_3=C_4(0)$&$\mathfrak{\mysosmall}(1,1;\frac\pi4) $&\\
IV & $A_5(0)$ & $\mathfrak{r}_3$ &Lsa&$C_5$&  $\mathfrak{a}_0$  &\\
V&  $A_4$ & $\mathfrak{r}_{3,1}$ &  La& $C_4(1)$&$\mathfrak{\mysosmall}(2;0)=\mathfrak{\mysosmall}(1,1;0)$ &\\
VI$_{h<0}$  & $A_5(\lambda<0)$ &$\mathfrak{r}_{3,\mu}$ & Lssa-($h$)&$C_4(\mu)$& $\mathfrak{\mysosmall}(1,1;\theta)$ &$\theta\in(0,\frac\pi2)$\\
VII$_{h>0}$   &  $A_5(\lambda>0)$ &$\mathfrak{r'}_{3,h}$ &Lssa+($h$)&$C_6(h)$& $\mathfrak{\mysosmall}(2;\theta)$& $\theta\in(0,\frac\pi2)$ \\
VIII &$A_8$&${\mathfrak{sl}_2}$& Lsss-&$C_8$&  ${\mathfrak{sl}_2}$&  \\
IX  & $A_9$&$\mathfrak{su}_2$& Lsss+&$C_7$& $\mathfrak{su}_2$ & \\
VI$_0$ & $A_6$ & $\mathfrak{r}_{3,-1}$& Lss-&$C_4(-1)$&$\mathfrak{\mysosmall}(1,1;\frac\pi2)$ & \\
VII$_0$ & $A_7$ & $\mathfrak{n}_2$&Lss+&$C_6(0)$&$\mathfrak{\mysosmall}(2;\frac\pi2)$ & \\
\bottomrule
\end{tabular}
\end{center}
\caption{Classification of three-dimensional Lie algebras. The Behr invariant $h$ is as in \protect\cite{EllisMacCallum}, with 
$h=\pm\cot^2\theta$ for  $\mathfrak{\mysosmall}(2;\theta)$ and $\mathfrak{\mysosmall}(1,1;\theta)$ respectively, while Levy-Nahas uses the invariant $\lambda=\pm\tan^2\theta$. The scaling parameter $\mu$ is related to the other parameters by $h=-(\frac{1+\mu}{1-\mu})^2$ and $\mu=\frac{\cos\theta-\sin\theta}{\cos\theta+\sin\theta}$ for $-1\leq\mu<1$.} 
\label{table:morebianchi}
\end{table}
\end{landscape}
\clearpage
}
Let us establish some of these isomorphisms. The algebra $\mathfrak{r}'_{3,h}$ ($h\geq0$) is defined in Fialowski et al.~\cite{fialowski} by $[l,m_1]=h m_1 +m_2$ and $[l,m_2]=-m_1+h m_2$. Setting $h=\cot\theta$ and rescaling $l$ by $\sin\theta$ shows that it is isomorphic to $\mathfrak{\mysosmall}(2;\theta)$ ($\theta\in(0,\frac\pi2]$). The algebra $\mathfrak{r}_{3,\mu}$ ($|\mu|<1$) is defined by $[l,x_1]=x_1$ and $[l,x_2]=\mu\, x_2$. This is isomorphic to $\mathfrak{\mysosmall}(1,1;\theta)$ with $\theta\in(0,\frac\pi2)$. To show this, note that the action of $l$ on the vector $m_i$ is diagonalizable,
\begin{equation}
[l,\begin{pmatrix}m_1 \\ m_2\end{pmatrix}]=
\begin{pmatrix}
\cos\theta &\sin\theta\\\sin\theta&\cos\theta
\end{pmatrix}\begin{pmatrix}
m_1 \\ m_2
\end{pmatrix}
\end{equation}
and the ratio of the eigenvalues is $\mu=\frac{\cos\theta-\sin\theta}{\cos\theta+\sin\theta}$. One can include the values $\mu=\pm1$ in $\mathfrak{r}_{3,\mu}$ which correspond to $\theta=0$ and $\theta=\frac\pi2$ in $\mathfrak{\mysosmall}(1,1;\theta)$. In particular, $\mathfrak{r}_{3,1}=\mathfrak{\mysosmall}(2;0)=\mathfrak{\mysosmall}(1,1;0)$.  The algebra $\mathfrak{r}_{3}$ is defined in \cite{fialowski} by $[r,x]=x$ and $[r,y]=y+x$, which is precisely $\mathfrak{a}_0$.

Levy-Nahas~\cite{levy-nahas:deform} uses the matrix $\rho$ defined by $C_{l i}{}^{j}= \rho^{jk}\epsilon_{l i k}$ and the completely antisymmetric tensor $\epsilon_{ijk}$. With this, $A_5(\lambda)$ and $A_4$ have $\rho$-matrices
\begin{equation}
 \begin{pmatrix}
  1&1&0\\-1&\lambda&0\\0&0&0
 \end{pmatrix}~\text{ and }
 \begin{pmatrix}
  0&1&0\\-1&0&0\\0&0&0
 \end{pmatrix}~.\quad
\end{equation}
A short calculation translates these as corresponding to $\mathfrak{a}_\lambda$ and $\mathfrak{\mysosmall}(2;0)$ respectively, where $\mathfrak{a}_\lambda$ was defined in section \ref{sec:3dLie}. Recall also that the families $\mathfrak{\mysosmall}(2;\theta)$ and $\mathfrak{\mysosmall}(1,1;\theta)$ with $\theta\in(0,\frac\pi2)$ can be embedded in the family $\mathfrak{a}_\lambda$. Their parameters are related (respectively for the two algebras) by $\lambda=\tan^2\theta$ and $\lambda=-\tan^2\theta$. Levy-Nahas calls $A_2$ the algebra defined by $\rho$-matrix 
\begin{equation}
 \begin{pmatrix}
  1&0&0\\0&0&0\\0&0&0
 \end{pmatrix}~,
\end{equation}
which corresponds to the bracket $[r,x]=-x$. It matches our definition of $\mathfrak{a}_\infty$ and the definition of $\mathfrak{n}_3$ in Fialowski et al. Finally, $A_5(-1)$ is called specifically $A_3$, but is none other than $\mathfrak{\mysosmall}(1,1;\frac\pi4)=\mathfrak{r}_{3,0}$. 

In summary, we have shown the isomorphisms
\begin{align}
 A_4&=\mathfrak{\mysosmall}(2;0)=\mathfrak{r}_{3,1} \\
 A_5(0)&=\mathfrak{a}_0 = \mathfrak{r}_{3} \\
 A_3&=\mathfrak{\mysosmall}(1,1;\frac\pi4)=\mathfrak{r}_{3,0}
\end{align}
There is a mismatch of these three isomorphisms with table 3 of \cite{fialowski}. We also observe that the graph in figure 2 of \cite{fialowski} does not descibe the deformations $\mathfrak{\mysosmall}(2;0)\famdeforms\mathfrak{\mysosmall}(2;\theta)$ and $\mathfrak{a}_0\famdeforms\mathfrak{\mysosmall}(2;\theta)$. In Levy-Nahas we point out a typo where he writes the $\rho$-matrix of A$_3 = \mathfrak{\mysosmall}(1,1;\frac\pi4)$. It should be
\begin{equation}
 \begin{pmatrix}
  1&1&0\\-1&-1&0\\0&0&0
 \end{pmatrix}~,
\end{equation}
with brackets $[r,x]=x+y$ and $[r,y]=x+y$. The derived algebra has indeed dimension one, and A$_3$ can be embedded in A$_5(\lambda)=\mathfrak{a}_\lambda$ at $\lambda=-1$.

\bibliography{defSF}
\bibliographystyle{utphys}
\end{document}